\documentclass{easychair}

\usepackage[utf8]{inputenc}
\usepackage{coqdoc}
\usepackage{amsmath,amssymb}
\usepackage{graphicx}
\usepackage{algorithm}
\usepackage{algorithmic}

\def\dattest#1#2{ #1~:\triangleright~#2}
\def\attest#1#2{#1~\vartriangleright~#2}
\def\dattestat#1#2#3{ #1~:\triangleright_{=#3}~#2}
\def\attestat#1#2#3{#1~\vartriangleright_{=#3}~#2}

\def\attestaft#1#2#3{#1~\vartriangleright_{>#3}~#2}
\def\dattestbef#1#2#3{ #1~:\triangleright_{<#3}~#2}
\def\attestbef#1#2#3{#1~\vartriangleright_{<#3}~#2}

\newcommand{\INDSTATE}[1][1]{\STATE\hspace{#1ex}}

\newcommand{\INDTWO}{\INDSTATE[3]}

\newcommand{\INDFOUR}{\INDSTATE[7]}

\title{Towards Secure and Trusted-by-Design Smart Contracts}

\date{}
\author{Zaynah~Dargaye\inst{} \and \"{O}nder~G\"{u}rcan\inst{} \and Florent~Kirchner\inst{} \and Sara~Tucci~Piergiovanni\inst{}}
\institute{CEA, LIST, Point Courrier 174, Gif-sur-Yvette, F-91191 France   \\
 \email{lea-zaynah.dargaye@cea.fr, onder.gurcan@cea.fr, florent.kirchner@cea.fr, sara.tucci@cea.fr} \\
}

\authorrunning{Z.~Dargaye, \"{O}.~G\"{u}rcan, F.~Kirchner and S.~Tucci~Piergiovanni }
\titlerunning{Cyberlogic for Smart Contracts}

\begin{document}
\maketitle
\begin{abstract}
Distributed immutable ledgers, or blockchains, allow the secure
digitization of evidential transactions without relying on a trusted
third-party. Evidential transactions involve the exchange of any form
of physical evidence, such as money, birth certificate, visas,
tickets, etc. Most of the time, evidential transactions occur in the
context of complex procedures, called evidential protocols, among
physical agents. The blockchain provides the mechanisms to
transfer evidence, while smart contracts allow encoding evidential
protocols on top of a blockchain. Smart contracts are indeed programs
executing within the blockchain in a decentralized and replicated
fashion.

As a smart contract obviates the need of trusted third-party and
runs on several machines anonymously, it is a high critical
program that has to be secure and trusted-by-design. While most of the
current smart contract languages focus on easy programmability, they
do not directly address the need of guaranteeing trust and
accountability of evidential protocols encoded as smart contracts.
\end{abstract}

\section{Introduction} {\label{sec:intro}}

The rise of immutable distributed ledgers, or {\it blockchains},
extends the {\it ``code is law''} vision to organizations and
corporations through the Decentralized Autonomous Organization ({\it
  DAO}) concept. Indeed, blockchains allow the secure digitization of
{\it evidential transactions} without relying on a trusted
third-party~\cite{SCdef} -- where evidential transactions involve the
exchange of any form of physical evidence, such as money, birth
certificate, visas, tickets, etc. Most of the time, evidential
transactions occur in the context of complex procedures encoded by
specific programs, called {\it smart contracts}, executing within the
blockchain in a decentralized and replicated fashion. A DAO, in
particular, is an organization run by smart contracts encoding rules
of governance. For those organization, promises in terms of
independence and savings are great advantages that smart contract can
provide them. Indeed, smart contract allows the removed of both the
trusted third-party and the need for repetitious (often manual)
recording of contracts.

The smart contract concept plays a main role in a DAO. However, there
is no official definition of smart contract. So far, any agreement has
been established regarding this concept.  Originally in 1994
in~\cite{SCdef}, Nick Szabo defined a smart contract as {\it a
  computerized transaction protocol that executes the terms of a
  contract}. The general objectives are to satisfy common contract
conditions (such as payment terms, liens, confidentiality, and even
enforcement) and minimize the need for trusted intermediaries. Related
economic goals include lowering fraud loss, arbitration and
enforcement costs, and other transaction costs. Nick Szabo believes
that a contractual clause implemented in software and benefiting from
cryptographical mechanisms would make the infringement of contracts
very expensive -- it is exactly this aspect that makes the contract
``smart". Nowadays, with the rise of blockchains, a smart contract designates
any piece of code running in the blockchain, losing
the legal flavor but keeping its immutability properties -- once in
the blockchain the smart contract cannot be changed becoming the
``law" for the community using it.

Recently Josh Stark, Head of the Operation \& Legal at LedgerLabs, in
the Ethereum community made an attempt to clarify these notions
identifying two kinds of smart contracts (see~\cite{JStark}): {\it
  smart legal contracts} and {\it smart contract codes}, So far, most
of the contribution has focused on improving and designing smart
contract code. Moreover, smart contract code languages, such as
Solidity for Ethereum, lack both formal foundations and expressiveness
to program smart legal contracts in a secure way. Plus, a large number
of vulnerabilities coming from the execution of the programs in a
widely distributed network, i.e., the blockchain network, are fully
exploitable today.

The lack of formalization and verification has already shown its
impact: the famous attack to {\it The DAO}~\cite{theDAO} was caused by
a simple bug in the smart contract. The survey~\cite{Atzei1} shows
that most attacks in Ethereum were caused by bugs or vulnerabilities
of the execution platform (Ethereum Virtual Machine and blockchain
network). Recently, the Tezos ledger has taken several steps to
address these problems, with the use of OCaml for its implementation,
and Michelson, a statically typed functional language for its smart
contracts. However, the overall challenge of smart contract security
remains unchanged. We advocate that to tackle this challenge there is
the need (i) of a formal language specifically designed to capture the
notion of trust (accountability, authentication, privacy) for the
secure implementation of smart legal contracts in software, (ii) to
capture and abstract the subtleties of the execution of smart
contracts in a possible wide distributed environment and (iii) to
provide full support for a compilation chain towards smart contract
code language.

In this paper, we focus on the design of the smart legal contract
language. We advocate that such challenge could be addressed by
extending so-called trust management frameworks to the
blockchain. Trust management frameworks are formal frameworks allowing
to formalize the specification of evidential protocols and verify
their correct implementation using a trusted-by-design approach.  In
particular, this paper presents a Cyberlogic Coq implementation that
gathers main features of being a smart legal contract. Cyberlogic is
a trust management frameworks, a first-order logic featuring an
authority algebra for specifying protocols and their
policies. Therefore, the Coq implementation enables to mechanically
specify and verify Cyberlogic protocols. Using an implementation in
the Coq platform allows to provide computable proof, hence, properties
of Cyberlogic protocol proved will come with such a data. As the Cyberlogic
implementation is a shallow embedded one, a smart legal
contract specifies in Cyberlogic can be extracted to an executable language.

First, we describe the state-of-the-art of specification of evidential
protocols. Then, we detail our Cyberlogic framework and its
implementation in Coq. Secondly, we state that an extension to
Cyberlogic is the perfect candidate to design and implement
trustworthy smart contracts. We corroborate our statement by
specifying a smart contract in Cyberlogic and we present its
implementation in a smart contract language. Finally, we explicit the
current trails that we are exploring to provide a framework to design
secure and trusted-by-design smart contracts.

\section{State-of-the-Art: Specifying Evidential Protocols}{\label{sec:cyber}}

Evidential protocols by essence are based on trust between entities
and particularly third-parties such as governments, banks or
insurances. In an evidential protocol, those kind of entities exchange
specific data, called evidences. An evidence is an official
certificates, license or visa. It constitutes critical data, and is
usually stated or claimed by a specific authority: government,
embassy, bank... Such authority has a role of trusted third-party, as
they represent a real world authority. Even though, the accountability is
endorsed by the authority which claims it, an evidence has to be
formally specified as well as another data in a protocol. Smart
contracts have a lot in common with evidential protocols: a smart
contract represents the interaction between the blockchain and the
real world. Plus, smart contracts implementing a DAO are
supposed to manage data that represent real world evidence: passport
number, or bank account or visa....

In this section, we detail the state-of-the-art of evidential
protocols specification. First, we present trust management frameworks
which allow to specify trust and accountability. Second, we focus
on the Cyberlogic framework which is more than a trust management
framework as it allows reasoning and verification. Finally,
we detail our Cyberlogic theorem prover in Coq. 

\subsection{Trust Management}

The underlying concepts of evidential protocols are the accountability
of the evidences,their establishment and its usage as well as the
trust in evidences.  Therefore, the specification of evidential
protocols requires being able to specify trust and accountability,
which is the purpose of \emph{trust management}.

The verification of authentication is a related domain of trust
management. For example, the PGP platform validates the link between
public keys and agents in decentralized web protocols and the X.509
platform deals with hierarchy in the same kind of protocols. In
other words, X.509 deals with authorities.

PolicyMaker \cite{Blaze98} is a trust management tool which associates
to the management of the security (via public/private keys) a logic of
predicates on keys. PolicyMaker is a language which features the
ability of inserting logic assertions. These assertions can express
that a certain key permits (authorizes) a certain predicate. However,
PolicyMaker does not allow formal reasoning for computing a proof to validate
the claim of an authority (abstraction of the entity that owns the
key). The descendant of PolicyMaker is
KeyNote~\cite{DBLP:conf/spw/BlazeFK98}, an other trust management
framework is Fidelis~\cite{Fidelis}. This approach tackles
specification of policies in evidential transactions.

One the other hand, some contribution has been done about verification of
trust in evidential protocols. In particular, a first alternative is
to use the {\it a posteriori} validation of the authentication
\cite{Appel1999}. The proof accompanies the data and is verified at
the reception by the authority. It has a cost and adds some work to
the authority. \cite{LGF00} describes a delegation logic which
considers the delegation of the authority. This logic permits
expressing relation such as: - A claims P, - A delegates P to B with a
delegation of k depth (a trusted chain of k authorities), - A claims
for B about P. Cyberlogic concentrates a large number of advantages of all
the previous tools.

\subsection{Cyberlogic}

The Cyberlogic framework inherits the major features of trust
management systems. In addition, Cyberlogic features native
constructions to deal with distributed systems.

Cyberlogic~\cite{RS:HCSS03} is designed for the transition from paper
documents to electronic documents. In particular, Cyberlogic
offers a framework for formalization and reasoning on attestation
elements such as visas, certificates \emph{etc}$\ldots$ Cyberlogic is
a logic and a distributed program in which protocols can play several
exchanges of attestations. As Cyberlogic features a first order logic,
it allows to reason upon the actions of a Cyberlogic protocol.
As Cyberlogic also features an authorities' algebra
it allows to reason upon trust and accountability.

In addition to the expression of certificates and visas, we believe that the
authorities' algebra can also be useful in complex systems
verification, and whenever trust into an entity takes the place
of a formal verification : acceptance.

Let's suppose a development using a library in a black box. The library
editor claims that the library has been formally verified. Then, the
user can suppose that each function of this library obeys to its
specification. In that configuration, which is not rare, the
Cyberlogic gives some feature to naturally express some hypotheses
usually categorized as informal hypotheses, or working hypotheses. If
$Spec$ is the specification of the function $f$ of the library, then the
user can use the Cyberlogic to formalize its development. Hence, the
specification of f which becomes a formal formula: $\attest{E}{Spec(f)}$. If the
editor has used Cyberlogic to formalize the library, then the library
specification can be reused by the user. Finally, if f does not fill
its specification and appears to bug in the development, the user
can identify the editor as guilty and accountable.

An interpreter for Cyberlogic protocols and formulas has been
developed at SRI International, in the PVS system~\cite{pvs}. Since
2013 at CEA List, thanks to two national projects (SystemX MIC and FUI
GeoTransMD) a theorem prover in Coq~\cite{Coq:manual} has been
developed and used to verify authenticity properties of distributed
protocols in communication systems.

In particular, the authors of~\cite{RS:HCSS03} states that Cyberlogic
is designed to enable evidential protocols implementation into
agent-based frameworks where agents execute Cyberlogic protocols to
carry out specific tasks that require the exchange of authorization
and authentication information. Nowadays, this definition seems quite
similar to blockchain protocols.

Therefore, Cyberlogic fits for formalizing complex systems. In particular,
we believe that Cyberlogic provides most of the properties required by
a smart legal contract language. 

\subsection{Shallow Embedding Implementation of Cyberlogic in Coq} 

 The following describes a formalization of the Cyberlogic language in
 a shallow embedding implementation in Coq. The aim of this
 formalization is to develop a prover for Cyberlogic formulas. The
 shallow embedding implementation allows to inherits all Gallina
 expressivness and to use the standard libraries of Coq.
 Moreover, it allows to focus only on the special
 formulas of Cyberlogic: the attestation. An attestation is a logic
 formulas which is claimed by an authority. The other Cyberlogic
 constructors are those of the Coq language itself.

\subsubsection{Authority and Attestations} 

 First, we define the notion of authority. Authorities are the owners
 of claims. A claim is a property of a data that is endorsed by an
 authority.  This authority is accountable for this
 property. An authority is the abstraction of a entity such as a
 person, a bank, a company or an organization. It can be any
 actors of a system as well, who can establish or claim a property.

  In Cyberlogic, an authority is uniquely identified and decidably
  comparable to other authorities. The type Authority is the
  abstraction of authorities.

\subsubsection{Different Kinds of Attestations} 
 An authority can claim a property either directly or
 indirectly. If an authority claims something directly then it
 could have establish it itself and is
 accountable for this claim. If an authority claims something
 indirectly, it means that another authority transmits her(him)
 this claim. In other words, the chain of trust can be trace to find the
 accountable authority. In the formalization we name the
 qualification direct or indirect by access. \begin{coqdoccode}
\coqdocemptyline
\coqdocnoindent
\coqdockw{Inductive} \coqdef{cyber.access}{access}{\coqdocinductive{access}} := \ensuremath{|} \coqdef{cyber.Direct}{Direct}{\coqdocconstructor{Direct}} : \coqref{cyber.access}{\coqdocinductive{access}} \ensuremath{|} \coqdef{cyber.Indirect}{Indirect}{\coqdocconstructor{Indirect}} : \coqref{cyber.access}{\coqdocinductive{access}}.\coqdoceol
\coqdocemptyline
\end{coqdoccode}
There are different kinds of claims (named {\tt authority} in the Coq
development). In addition to the combination of an authority and an access
mode, time can be take in account in some claims. In fact, as in modal
logics, it is possible to claim a fact before, at or after a
date. Then, a claim defines an owner, an access and a timing for a
claim. \begin{coqdoccode} \coqdocemptyline \coqdocnoindent
  \coqdockw{Inductive}
  \coqdef{cyber.authority}{authority}{\coqdocinductive{authority}}
  :=\coqdoceol \coqdocnoindent \ensuremath{|}
  \coqdef{cyber.Key}{Key}{\coqdocconstructor{Key}} :
  \coqref{cyber.Authority}{\coqdocaxiom{Authority}}
  \coqexternalref{:type scope:x '->'
    x}{http://coq.inria.fr/distrib/8.5pl3/stdlib/Coq.Init.Logic}{\coqdocnotation{\ensuremath{\rightarrow}}}
  \coqref{cyber.access}{\coqdocinductive{access}}
  \coqexternalref{:type scope:x '->'
    x}{http://coq.inria.fr/distrib/8.5pl3/stdlib/Coq.Init.Logic}{\coqdocnotation{\ensuremath{\rightarrow}}}
  \coqref{cyber.authority}{\coqdocinductive{authority}} \coqdoceol
  \coqdocnoindent \ensuremath{|}
  \coqdef{cyber.KatT}{KatT}{\coqdocconstructor{KatT}} :
  \coqref{cyber.Authority}{\coqdocaxiom{Authority}}
  \coqexternalref{:type scope:x '->'
    x}{http://coq.inria.fr/distrib/8.5pl3/stdlib/Coq.Init.Logic}{\coqdocnotation{\ensuremath{\rightarrow}}}
  \coqref{cyber.access}{\coqdocinductive{access}}
  \coqexternalref{:type scope:x '->'
    x}{http://coq.inria.fr/distrib/8.5pl3/stdlib/Coq.Init.Logic}{\coqdocnotation{\ensuremath{\rightarrow}}}
  \coqref{cyber.time}{\coqdocdefinition{time}} \coqexternalref{:type
    scope:x '->'
    x}{http://coq.inria.fr/distrib/8.5pl3/stdlib/Coq.Init.Logic}{\coqdocnotation{\ensuremath{\rightarrow}}}
  \coqref{cyber.authority}{\coqdocinductive{authority}}\coqdoceol
  \coqdocnoindent \ensuremath{|}
  \coqdef{cyber.Kbef}{Kbef}{\coqdocconstructor{Kbef}} :
  \coqref{cyber.Authority}{\coqdocaxiom{Authority}}
  \coqexternalref{:type scope:x '->'
    x}{http://coq.inria.fr/distrib/8.5pl3/stdlib/Coq.Init.Logic}{\coqdocnotation{\ensuremath{\rightarrow}}}
  \coqref{cyber.access}{\coqdocinductive{access}}
  \coqexternalref{:type scope:x '->'
    x}{http://coq.inria.fr/distrib/8.5pl3/stdlib/Coq.Init.Logic}{\coqdocnotation{\ensuremath{\rightarrow}}}
  \coqref{cyber.time}{\coqdocdefinition{time}} \coqexternalref{:type
    scope:x '->'
    x}{http://coq.inria.fr/distrib/8.5pl3/stdlib/Coq.Init.Logic}{\coqdocnotation{\ensuremath{\rightarrow}}}
  \coqref{cyber.authority}{\coqdocinductive{authority}}\coqdoceol
  \coqdocnoindent \ensuremath{|}
  \coqdef{cyber.Kaft}{Kaft}{\coqdocconstructor{Kaft}} :
  \coqref{cyber.Authority}{\coqdocaxiom{Authority}}
  \coqexternalref{:type scope:x '->'
    x}{http://coq.inria.fr/distrib/8.5pl3/stdlib/Coq.Init.Logic}{\coqdocnotation{\ensuremath{\rightarrow}}}
  \coqref{cyber.access}{\coqdocinductive{access}}
  \coqexternalref{:type scope:x '->'
    x}{http://coq.inria.fr/distrib/8.5pl3/stdlib/Coq.Init.Logic}{\coqdocnotation{\ensuremath{\rightarrow}}}
  \coqref{cyber.time}{\coqdocdefinition{time}} \coqexternalref{:type
    scope:x '->'
    x}{http://coq.inria.fr/distrib/8.5pl3/stdlib/Coq.Init.Logic}{\coqdocnotation{\ensuremath{\rightarrow}}}
  \coqref{cyber.authority}{\coqdocinductive{authority}}.\coqdoceol
  \coqdocemptyline
\end{coqdoccode}
\subsubsection{Attestations}

 Finally, an attestation is a property which is claimed by a authority. \begin{coqdoccode}
\coqdocemptyline
\coqdocnoindent
\coqdockw{Variable} \coqdef{cyber.attestation}{attestation}{\coqdocvariable{attestation}} : \coqref{cyber.authority}{\coqdocinductive{authority}} \coqexternalref{:type scope:x '->' x}{http://coq.inria.fr/distrib/8.5pl3/stdlib/Coq.Init.Logic}{\coqdocnotation{\ensuremath{\rightarrow}}} \coqdockw{Prop} \coqexternalref{:type scope:x '->' x}{http://coq.inria.fr/distrib/8.5pl3/stdlib/Coq.Init.Logic}{\coqdocnotation{\ensuremath{\rightarrow}}} \coqdockw{Prop}.\coqdoceol
\coqdocemptyline
\end{coqdoccode}
For the sake of clarity, let's define some notations as similar as possible 
  from those of the original paper~\cite{CyberL}. \\
  \begin{small}
  \begin{tabular}{llr}
  $k~|>~f$ & $ \attest{k}{f}$ & $k$ attests directly $f$\\
  $k~*|>~f$ & $ \dattest{k}{f}$ & $k$ attests indirectly $f$ \\
  $k~|>=t~f $ & $\attestat{k}{f}{t}$ & $k$ attests directly $f$ at $t$ \\
  $k~*|>=t~f $ & $\dattestat{k}{f}{t}$ & $k$ attests indirectly $f$ at $t$ \\
  $k~|><t~f$ & $\attestbef{k}{f}{t} $ & $k$ attests directly $f$ before $t$ \\ 
  $k~*|><t~f$ & $\dattestbef{k}{f}{t} $ & $k$ attests indirectly $f$ before $t$ \\ 
  $k~|>>~t~f$ & $\attestaft{k}{f}{t}$ & $k$ attests directly $f$ after $t$ \\
  \end{tabular}
  \end{small}

\subsubsection{Delegation of Attestation} 

 Cyberlogic enables to formalize delegation of authority using the
 indirect mode of access. In fact, the indirect mode of access is
 obtained by a delegation from an authority to another. D1 is the
 direct delegation operator. D1(A,B,P) says that A claims P and A
 knows P from B, B is accountable for P. D2(A,B,P) says that A has
 been directly delegated by some other authority (C) to claim P. But C
 is not accountable for P, B is. D2 is a special delegation chain of
 depth 3. A delegation is indexed by the length of the chain of
 authority from the original claimer to the authority which actually
 claims.

  \begin{coqdoccode} \coqdocemptyline \coqdocnoindent
   \coqdockw{Definition}
   \coqdef{cyber.Dinf}{Dinf}{\coqdocdefinition{Dinf}} (\coqdocvar{k}
   \coqdocvar{k'}: \coqref{cyber.Authority}{\coqdocaxiom{Authority}})
   (\coqdocvar{A}:\coqdockw{Prop}) := \coqexternalref{:type scope:x
     '->'
     x}{http://coq.inria.fr/distrib/8.5pl3/stdlib/Coq.Init.Logic}{\coqdocnotation{(}}\coqdocvariable{k'}
   \coqref{cyber.::x '*|>' x}{\coqdocnotation{*|>}}
   \coqdocvariable{A}\coqexternalref{:type scope:x '->'
     x}{http://coq.inria.fr/distrib/8.5pl3/stdlib/Coq.Init.Logic}{\coqdocnotation{)->}}
   \coqexternalref{:type scope:x '->'
     x}{http://coq.inria.fr/distrib/8.5pl3/stdlib/Coq.Init.Logic}{\coqdocnotation{(}}\coqdocvariable{k}
   \coqref{cyber.::x '*|>' x}{\coqdocnotation{*|>}}
   \coqdocvariable{A}\coqexternalref{:type scope:x '->'
     x}{http://coq.inria.fr/distrib/8.5pl3/stdlib/Coq.Init.Logic}{\coqdocnotation{)}}.\coqdoceol
   \coqdocemptyline \coqdocnoindent \coqdockw{Definition}
   \coqdef{cyber.D1}{D1}{\coqdocdefinition{D1}} (\coqdocvar{k}
   \coqdocvar{k'}: \coqref{cyber.Authority}{\coqdocaxiom{Authority}})
   (\coqdocvar{A}:\coqdockw{Prop}):= \coqexternalref{:type scope:x
     '->'
     x}{http://coq.inria.fr/distrib/8.5pl3/stdlib/Coq.Init.Logic}{\coqdocnotation{(}}\coqdocvariable{k'}\coqref{cyber.::x
     '|>'
     x}{\coqdocnotation{|>}}\coqdocvariable{A}\coqexternalref{:type
     scope:x '->'
     x}{http://coq.inria.fr/distrib/8.5pl3/stdlib/Coq.Init.Logic}{\coqdocnotation{)}}
   \coqexternalref{:type scope:x '->'
     x}{http://coq.inria.fr/distrib/8.5pl3/stdlib/Coq.Init.Logic}{\coqdocnotation{\ensuremath{\rightarrow}}}
   \coqexternalref{:type scope:x '->'
     x}{http://coq.inria.fr/distrib/8.5pl3/stdlib/Coq.Init.Logic}{\coqdocnotation{(}}\coqdocvariable{k}
   \coqref{cyber.::x '*|>' x}{\coqdocnotation{*|>}}
   \coqdocvariable{A}\coqexternalref{:type scope:x '->'
     x}{http://coq.inria.fr/distrib/8.5pl3/stdlib/Coq.Init.Logic}{\coqdocnotation{)}}.\coqdoceol
   \coqdocemptyline \coqdocnoindent \coqdockw{Definition}
   \coqdef{cyber.D2}{D2}{\coqdocdefinition{D2}} (\coqdocvar{k}
   \coqdocvar{k'}: \coqref{cyber.Authority}{\coqdocaxiom{Authority}})
   (\coqdocvar{A}:\coqdockw{Prop}):= \coqdoceol \coqdocindent{0.50em}
   \coqexternalref{:type scope:x '->'
     x}{http://coq.inria.fr/distrib/8.5pl3/stdlib/Coq.Init.Logic}{\coqdocnotation{(}}\coqdockw{\ensuremath{\forall}}
   (\coqdocvar{k0} :
   \coqref{cyber.Authority}{\coqdocaxiom{Authority}}),
   \coqexternalref{:type scope:x '/x5C'
     x}{http://coq.inria.fr/distrib/8.5pl3/stdlib/Coq.Init.Logic}{\coqdocnotation{(}}\coqdocvariable{k0}
   \coqref{cyber.::x '|>' x}{\coqdocnotation{|>}}
   \coqdocvariable{A}\coqexternalref{:type scope:x '/x5C'
     x}{http://coq.inria.fr/distrib/8.5pl3/stdlib/Coq.Init.Logic}{\coqdocnotation{)}}
   \coqexternalref{:type scope:x '/x5C'
     x}{http://coq.inria.fr/distrib/8.5pl3/stdlib/Coq.Init.Logic}{\coqdocnotation{\ensuremath{\land}}}
   \coqexternalref{:type scope:x '/x5C'
     x}{http://coq.inria.fr/distrib/8.5pl3/stdlib/Coq.Init.Logic}{\coqdocnotation{(}}\coqdocvariable{k'}
   \coqref{cyber.::x '|>'
     x}{\coqdocnotation{|>(}}\coqexternalref{:type scope:x '->'
     x}{http://coq.inria.fr/distrib/8.5pl3/stdlib/Coq.Init.Logic}{\coqdocnotation{(}}\coqdocvariable{k0}
   \coqref{cyber.::x '|>' x}{\coqdocnotation{|>}}
   \coqdocvariable{A}\coqexternalref{:type scope:x '->'
     x}{http://coq.inria.fr/distrib/8.5pl3/stdlib/Coq.Init.Logic}{\coqdocnotation{)->}}\coqdocvariable{A}\coqref{cyber.::x
     '|>' x}{\coqdocnotation{)}}\coqexternalref{:type scope:x '/x5C'
     x}{http://coq.inria.fr/distrib/8.5pl3/stdlib/Coq.Init.Logic}{\coqdocnotation{)}}\coqexternalref{:type
     scope:x '->'
     x}{http://coq.inria.fr/distrib/8.5pl3/stdlib/Coq.Init.Logic}{\coqdocnotation{)}}
   \coqexternalref{:type scope:x '->'
     x}{http://coq.inria.fr/distrib/8.5pl3/stdlib/Coq.Init.Logic}{\coqdocnotation{\ensuremath{\rightarrow}}}
   \coqexternalref{:type scope:x '->'
     x}{http://coq.inria.fr/distrib/8.5pl3/stdlib/Coq.Init.Logic}{\coqdocnotation{(}}\coqdocvariable{k}
   \coqref{cyber.::x '*|>' x}{\coqdocnotation{*|>}}
   \coqdocvariable{A}\coqexternalref{:type scope:x '->'
     x}{http://coq.inria.fr/distrib/8.5pl3/stdlib/Coq.Init.Logic}{\coqdocnotation{)}}.\coqdoceol
\end{coqdoccode}
\subsubsection{Time Reasoning in the Cyberlogic} 

One other interesting and useful feature of Cyberlogic is the modal
logic integration in the authority algebra. Moreover, reasoning about
time is facilitated thanks to a specific authority accountable for time:
$Kt$. $Kt$ establishes the current time and characterizes if a date is in
past or in future.

\begin{coqdoccode}
\coqdocemptyline
\coqdocnoindent
\coqdockw{Variable} \coqdef{cyber.hasbeen}{hasbeen}{\coqdocvariable{hasbeen}}: \coqref{cyber.time}{\coqdocdefinition{time}} \coqexternalref{:type scope:x '->' x}{http://coq.inria.fr/distrib/8.5pl3/stdlib/Coq.Init.Logic}{\coqdocnotation{\ensuremath{\rightarrow}}}\coqdockw{ Prop}.\coqdoceol
\coqdocemptyline
\coqdocemptyline
\coqdocnoindent
\coqdockw{Definition} \coqdef{cyber.is time}{is\_time}{\coqdocdefinition{is\_time}} (\coqdocvar{t}:\coqref{cyber.time}{\coqdocdefinition{time}}):\coqdockw{Prop}:= \coqref{cyber.Kt}{\coqdocaxiom{Kt}} \coqref{cyber.::x '*|>' x}{\coqdocnotation{*|>}} \coqref{cyber.::x '*|>' x}{\coqdocnotation{(}}\coqref{cyber.hasbeen}{\coqdocaxiom{hasbeen}} \coqdocvariable{t}\coqref{cyber.::x '*|>' x}{\coqdocnotation{)}}.\coqdoceol
\coqdocemptyline
\end{coqdoccode}
Definition of the current time. \begin{coqdoccode}
\coqdocemptyline
\coqdocnoindent
\coqdockw{Definition} \coqdef{cyber.curr}{curr}{\coqdocdefinition{curr}} (\coqdocvar{t}:\coqref{cyber.time}{\coqdocdefinition{time}}):\coqdockw{Prop}:= \coqref{cyber.is time}{\coqdocdefinition{is\_time}} \coqdocvariable{t} \coqexternalref{:type scope:x '/x5C' x}{http://coq.inria.fr/distrib/8.5pl3/stdlib/Coq.Init.Logic}{\coqdocnotation{\ensuremath{\land}}} \coqexternalref{:type scope:x '/x5C' x}{http://coq.inria.fr/distrib/8.5pl3/stdlib/Coq.Init.Logic}{\coqdocnotation{(}}\coqdockw{\ensuremath{\forall}} (\coqdocvar{t'}:\coqref{cyber.time}{\coqdocdefinition{time}}), \coqexternalref{:type scope:x '->' x}{http://coq.inria.fr/distrib/8.5pl3/stdlib/Coq.Init.Logic}{\coqdocnotation{(}}\coqref{cyber.is time}{\coqdocdefinition{is\_time}} \coqdocvariable{t'}\coqexternalref{:type scope:x '->' x}{http://coq.inria.fr/distrib/8.5pl3/stdlib/Coq.Init.Logic}{\coqdocnotation{)}} \coqexternalref{:type scope:x '->' x}{http://coq.inria.fr/distrib/8.5pl3/stdlib/Coq.Init.Logic}{\coqdocnotation{\ensuremath{\rightarrow}}} \coqdocvariable{t'}\coqexternalref{:nat scope:x '<=' x}{http://coq.inria.fr/distrib/8.5pl3/stdlib/Coq.Init.Peano}{\coqdocnotation{\ensuremath{\le}}} \coqdocvariable{t} \coqexternalref{:type scope:x '/x5C' x}{http://coq.inria.fr/distrib/8.5pl3/stdlib/Coq.Init.Logic}{\coqdocnotation{)}}.\coqdoceol
\end{coqdoccode}
Defining that a time is in future.  \begin{coqdoccode}
\coqdocemptyline
\coqdocnoindent
\coqdockw{Definition} \coqdef{cyber.in future}{in\_future}{\coqdocdefinition{in\_future}} (\coqdocvar{t}:\coqref{cyber.time}{\coqdocdefinition{time}}):= \coqref{cyber.Kt}{\coqdocaxiom{Kt}} \coqref{cyber.::x '*|>' x}{\coqdocnotation{*|>}} \coqexternalref{:type scope:'x7E' x}{http://coq.inria.fr/distrib/8.5pl3/stdlib/Coq.Init.Logic}{\coqdocnotation{\~{}(}}\coqref{cyber.hasbeen}{\coqdocaxiom{hasbeen}} \coqdocvariable{t}\coqexternalref{:type scope:'x7E' x}{http://coq.inria.fr/distrib/8.5pl3/stdlib/Coq.Init.Logic}{\coqdocnotation{)}}.\coqdoceol
\end{coqdoccode}

\section{Cyberlogic for Specifying Smart Legal Contracts}{\label{sec:schengen}}

In Section~\ref{sec:cyber}, we have shown how Cyberlogic allows reasoning
both on (i) actions, thanks to a first-order logic and (ii) on trust
and accountability, thanks to the authority algebra. Being
specifically designed for evidential protocols and benefiting from a
theorem prover in Coq, we advocate that Cyberlogic is, indeed, a perfect
candidate to be at the core of a smart legal contract language. To
support our claim, in this section, we show Cyberlogic at work by
specifying the Schengen visa management process as a Cyberlogic
protocol and present the corresponding smart contract code in
Solidity\footnote{The availability of a formal and proved
  specification opens the way for a formal compilation chain targeting
  the smart contract code; the compilation chain, however, is not in
  the scope of this paper (see section~\ref{sec:futurs} for a detailed
  future work discussion.}.

\subsection{Schengen Visa Management}

The Schengen visa management which is describe in accordance with
\href{https://ec.europa.eu/home-affairs/what-we-do/policies/borders-and-visas/visa-policy/required_documents_en}{
  the European Commission}, is an evidential protocol
where the visa represents an evidence delivered by a country of the
Schengen area. In order to obtain that evidence, it also requires to
gather other evidence (in accordance to the reuirements of the visa
management process). The visa itself is an evidence stating that all the
requirements are satisfied.

Nowadays, the execution of visa management protocol is time consuming
(latency experienced for each required evidence) and mostly manual
(photocopies, fulfilling forms, subforms...). The digitalisation of
the entire protocol through smart legal contracts is indeed a good
opportunity to improve the protocol and make it more secure. In the
reminder of this section we provide a characterisation of the visa
management as a Cyberlogic protocol, highlighting the kind of
reasoning that can be made. In particular, the Cyberlogic
specification allows the verifier of the visa (e.g. the custom's
officer that during a control spots a problem) to roll-up the entire
chain of trust to find the authority accountable for the possible
error.

\subsection{The Shengen Visa as a Cyberlogic Protocol}{\label{sec:alg}}

We propose to implement the Schengen visa management protocol as a
smart legal contract.  The result would be to alleviate centralization
and possible related bottlenecks by digitizing the process in a secure
and decentralized way \footnote{We are aware that the adoption of such
  a smart legal contract is not only a technical issue, since legal
  compliance with the proposed framework should also be
  evaluated. However, since our goal is to show how CyberLogic works,
  we are not taking into consideration legal issues. }. Let us note,
indeed, that the intermediary here is the smart legal contract itself
that will be later encoded as a smart contract code running in a
blockchain. The smart contract code will, indeed, (i) encode all the
rules pertaining to the management of the Schengen's visa, including
rules on authorities that have the right to deliver it, and (ii) will
be executed in a secured and decentralised fashion among the
blockchain participants.

Let us now to explicit our smart legal contract. First of all, to
satisfy the autonomy and the authority of the real states of the
Schengen area, the smart legal contract includes a consulate (or
prefecture) role as official authority having the rights to deliver
the visa. At delivery time, this official authority delegates the visa
to the requester, where: (i) but the official authority is still the
unique accountable for the visa and (ii) and the requester is allowed
to provide directly the required evidences.

Let us note that the consulate can be viewed as a trusted third party
we are still relying on. On the other hand, we propose to formally
delegate the entire procedure to the requester and to exploit
blockchain immutability properties to correctly compute proofs on
executed scenarios. This approach without fully realising the vision
of a decentralised and autonomous organisation, represents a first
step in this direction \footnote{Our approach is still compliant with
  the actual organisation of States and their rules, a different fully
  decentralized organisation would be possible using
  self-certification and new governance rules.}.  Anyway, a smart
legal contract to manage the Schengen visa will allow active controls,
transparency and detection of contradictories requirements.  A
controler will be able to go behind the visa itself and consult the
requirement it reprensent. In case of two requirements are
contradictories, the controller will be able to observe it.

The management of the Schengen visa as a smart legal contract in
Cyberlogic is composed of 4 functions which are: its demand, its
delivery, its control and its indictment. Let's informally define
these 4 functions.
\\*

{\bf To demand} a Schengen visa is to write evidential requirements of
the Visa's protocol in the blockchain, once and for all thanks to
immutability.  The accountability is preserved thanks to the
Cyberlogic protocol, where each evidential requirement is a claim from
the appropriate authority. \\* {\bf To deliver} a Schengen visa is to
write the visa in the blockchain, the visa is claimed by the consulate
(or official state organization) and is delegated to the requester (as
discussed above), thanks to the Cyberlogic delegation.  The chain of
trust from the visa is digitized thanks to the delegation mechanism of
Cyberlogic. \\* {\bf To control} a Schengen visa is a read in the
blockchain, accessing to the evidential requirements by rolling-up the
trust chain. \\* {\bf To suspect} a Schengen visa is to extract the
evidential requirements from the blockchain, analysing the evidential
requirements and identifying the potential suspicious claims, i.e.
computing accountability.

\subsection{The Cyberlogic Protocol}
  In the following we present the Cyberlogic protocol in the theorem prover of Cyberlogic.

As we define a smart contract, the code will exchange transactions that
  carry a demand, a visa or a control.

\begin{coqdoccode}
\coqdocemptyline
\coqdocnoindent
\coqdockw{Inductive} \coqdef{evidential.transaction}{transaction}{\coqdocaxiom{transaction}}:=\coqdoceol
\coqdocnoindent
\ensuremath{|} \coqdef{evidential.Demand}{Demand}{\coqdocconstructor{Demand}}: \coqdocaxiom{Schengen\_demand} \coqexternalref{:type scope:x '->' x}{http://coq.inria.fr/distrib/8.5pl3/stdlib/Coq.Init.Logic}{\coqdocnotation{\ensuremath{\rightarrow}}} \coqref{evidential.transaction}{\coqdocaxiom{transaction}}\coqdoceol
\coqdocnoindent
\ensuremath{|} \coqdef{evidential.Deliver}{Deliver}{\coqdocconstructor{Deliver}} : \coqdocaxiom{visa} \coqexternalref{:type scope:x '->' x}{http://coq.inria.fr/distrib/8.5pl3/stdlib/Coq.Init.Logic}{\coqdocnotation{\ensuremath{\rightarrow}}} \coqref{evidential.transaction}{\coqdocaxiom{transaction}}\coqdoceol
\coqdocnoindent
\ensuremath{|} \coqdef{evidential.Control}{Control}{\coqdocconstructor{Control}} : \coqdocaxiom{visa} \coqexternalref{:type scope:x '->' x}{http://coq.inria.fr/distrib/8.5pl3/stdlib/Coq.Init.Logic}{\coqdocnotation{\ensuremath{\rightarrow}}} \coqref{evidential.transaction}{\coqdocaxiom{transaction}}.\coqdoceol
\coqdocemptyline
\end{coqdoccode}
Queries are the suspicions that an officer can make. When an officer suspects a visa, he will make queries about it. \begin{coqdoccode}
\coqdocemptyline
\coqdocnoindent
\coqdockw{Definition} \coqdef{evidential.query}{query}{\coqdocdefinition{query}} := \coqdocaxiom{visa} \coqexternalref{:type scope:x '->' x}{http://coq.inria.fr/distrib/8.5pl3/stdlib/Coq.Init.Logic}{\coqdocnotation{\ensuremath{\rightarrow}}} \coqdockw{Prop}.\coqdoceol
\coqdocnoindent
\coqdockw{Definition} \coqdef{evidential.queries}{queries}{\coqdocdefinition{queries}} := \coqexternalref{list}{http://coq.inria.fr/distrib/8.5pl3/stdlib/Coq.Init.Datatypes}{\coqdocaxiom{list}} \coqref{evidential.query}{\coqdocdefinition{query}}.\coqdoceol
\coqdocemptyline
\coqdocemptyline
\end{coqdoccode}
Upon verification of an evidence, either the smart contract answers that 
  the visa is valid
  or there is a list of suspicious claims about the visa. \begin{coqdoccode}
\coqdocemptyline
\coqdocnoindent
\coqdockw{Inductive} \coqdef{evidential.answer}{answer}{\coqdocaxiom{answer}}:=\coqdoceol
\coqdocnoindent
\ensuremath{|} \coqdef{evidential.Valid}{Valid}{\coqdocconstructor{Valid}} : \coqref{evidential.answer}{\coqdocaxiom{answer}}\coqdoceol
\coqdocnoindent
\ensuremath{|} \coqdef{evidential.Suspects}{Suspects}{\coqdocconstructor{Suspects}} : \coqexternalref{list}{http://coq.inria.fr/distrib/8.5pl3/stdlib/Coq.Init.Datatypes}{\coqdocaxiom{list}} \coqdockw{Prop} \coqexternalref{:type scope:x '->' x}{http://coq.inria.fr/distrib/8.5pl3/stdlib/Coq.Init.Logic}{\coqdocnotation{\ensuremath{\rightarrow}}} \coqref{evidential.answer}{\coqdocaxiom{answer}}.\coqdoceol
\coqdocemptyline
\coqdocemptyline
\end{coqdoccode}
Action designates the actions that interact with the blockchain.
The interaction with the blockchain consists in three operations on transactions:
reading the ledger, writing in the ledger and questioning the ledger about properties
on a specific data. The API of the interaction  presented  here is quite light but sufficient
for highlighting the Cyberlogic properties.

\begin{coqdoccode}

\coqdocemptyline
\coqdocnoindent
\coqdockw{Variable} \coqdef{evidential.action}{action}{\coqdocvariable{action}}:\coqdockw{Type}.\coqdoceol
\coqdocemptyline
\coqdocnoindent
\coqdockw{Variable} \coqdef{evidential.write}{write}{\coqdocvariable{write}}: \coqdocaxiom{Authority} \coqexternalref{:type scope:x '->' x}{http://coq.inria.fr/distrib/8.5pl3/stdlib/Coq.Init.Logic}{\coqdocnotation{\ensuremath{\rightarrow}}} \coqref{evidential.transaction}{\coqdocaxiom{transaction}} \coqexternalref{:type scope:x '->' x}{http://coq.inria.fr/distrib/8.5pl3/stdlib/Coq.Init.Logic}{\coqdocnotation{\ensuremath{\rightarrow}}} \coqref{evidential.action}{\coqdocaxiom{action}}.\coqdoceol
\coqdocemptyline
\coqdocnoindent
\coqdockw{Variable} \coqdef{evidential.read}{read}{\coqdocvariable{read}} : \coqdocaxiom{Authority} \coqexternalref{:type scope:x '->' x}{http://coq.inria.fr/distrib/8.5pl3/stdlib/Coq.Init.Logic}{\coqdocnotation{\ensuremath{\rightarrow}}} \coqref{evidential.transaction}{\coqdocaxiom{transaction}} \coqexternalref{:type scope:x '->' x}{http://coq.inria.fr/distrib/8.5pl3/stdlib/Coq.Init.Logic}{\coqdocnotation{\ensuremath{\rightarrow}}} \coqref{evidential.action}{\coqdocaxiom{action}}.\coqdoceol
\coqdocemptyline
\coqdocnoindent
\coqdockw{Variable} \coqdef{evidential.verify}{verify}{\coqdocvariable{verify}} : \coqdocaxiom{Authority} \coqexternalref{:type scope:x '->' x}{http://coq.inria.fr/distrib/8.5pl3/stdlib/Coq.Init.Logic}{\coqdocnotation{\ensuremath{\rightarrow}}} \coqdocaxiom{visa} \coqexternalref{:type scope:x '->' x}{http://coq.inria.fr/distrib/8.5pl3/stdlib/Coq.Init.Logic}{\coqdocnotation{\ensuremath{\rightarrow}}} \coqref{evidential.queries}{\coqdocdefinition{queries}} \coqexternalref{:type scope:x '->' x}{http://coq.inria.fr/distrib/8.5pl3/stdlib/Coq.Init.Logic}{\coqdocnotation{\ensuremath{\rightarrow}}} \coqref{evidential.answer}{\coqdocaxiom{answer}}.\coqdoceol
\coqdocemptyline
\end{coqdoccode}
The following are the functions of the Schengen visa smart contract. 
\begin{coqdoccode}
\coqdocemptyline
\coqdocnoindent
\coqdockw{Definition} \coqdef{evidential.demand}{demand}{\coqdocdefinition{demand}} (\coqdocvar{Requester}: \coqdocaxiom{Authority})(\coqdocvar{C}:\coqdocdefinition{country}) (\coqdocvar{f}:\coqdocaxiom{schengen\_form}) (\coqdocvar{pic}:\coqdocaxiom{photo})\coqdoceol
\coqdocnoindent
(\coqdocvar{pport}:\coqdocaxiom{passport})(\coqdocvar{trvls}:\coqdocdefinition{travel\_itinerary})(\coqdocvar{ins\_policy}:\coqdocaxiom{travel\_health})(\coqdocvar{accs}:\coqdocdefinition{accommodations})\coqdoceol
\coqdocnoindent
(\coqdocvar{suff}:\coqdocaxiom{sufficient\_means})(\coqdocvar{t}:\coqdocdefinition{time}):=\coqdoceol
\coqdocindent{1.50em}
\coqref{evidential.write}{\coqdocaxiom{write}} \coqdocvariable{Requester} (\coqref{evidential.Demand}{\coqdocconstructor{Demand}} (\coqdocconstructor{mkDemand} \coqdocvariable{f} \coqdocvariable{pic} \coqdocvariable{pport} \coqdocvariable{trvls} \coqdocvariable{ins\_policy} \coqdocvariable{accs} \coqdocvariable{suff} \coqdocvariable{t})).\coqdoceol
\coqdocemptyline
\coqdocnoindent
\coqdockw{Definition} \coqdef{evidential.deliver}{deliver}{\coqdocdefinition{deliver}} (\coqdocvar{cons}:\coqdocaxiom{Authority})(\coqdocvar{v}:\coqdocaxiom{visa}):= \coqref{evidential.write}{\coqdocaxiom{write}} \coqdocvariable{cons} (\coqref{evidential.Deliver}{\coqdocconstructor{Deliver}} \coqdocvariable{v}).\coqdoceol
\coqdocemptyline
\coqdocnoindent
\coqdockw{Definition} \coqdef{evidential.control}{control}{\coqdocdefinition{control}} (\coqdocvar{officer}:\coqdocaxiom{Authority})(\coqdocvar{v}:\coqdocaxiom{visa}):= \coqref{evidential.read}{\coqdocaxiom{read}} \coqdocvariable{officer} (\coqref{evidential.Control}{\coqdocconstructor{Control}} \coqdocvariable{v}).\coqdoceol
\coqdocemptyline
\coqdocnoindent
\coqdockw{Definition} \coqdef{evidential.suspect}{suspect}{\coqdocdefinition{suspect}} (\coqdocvar{officer}:\coqdocaxiom{Authority})(\coqdocvar{v}:\coqdocaxiom{visa}) (\coqdocvar{ev}:\coqref{evidential.queries}{\coqdocdefinition{queries}}):= \coqref{evidential.verify}{\coqdocaxiom{verify}} \coqdocvariable{officer} \coqdocvariable{v} \coqdocvariable{ev}.\coqdoceol
\coqdocemptyline
\end{coqdoccode}
\subsection{The Protocol Policies} \begin{coqdoccode}
\coqdocemptyline
\end{coqdoccode}
The policy of demanding a visa consists in the time stamping verification. 
  The requester is accountable of the demand action. \begin{coqdoccode}
\coqdocemptyline
\coqdocnoindent
\coqdockw{Definition} \coqdef{evidential.demanding}{demanding}{\coqdocdefinition{demanding}} (\coqdocvar{r}:\coqdocaxiom{Authority}) (\coqdocvar{t}:\coqdocdefinition{time}) (\coqdocvar{c}:\coqdocdefinition{country})
(\coqdocvar{d}:\coqdocaxiom{Schengen\_demand}):= \coqdoceol
\coqdocindent{1.00em}
\coqdocprojection{time\_stamp} \coqdocvariable{d} \coqexternalref{:type scope:x '=' x}{http://coq.inria.fr/distrib/8.5pl3/stdlib/Coq.Init.Logic}{\coqdocnotation{=}} \coqdocvariable{t} \coqexternalref{:type scope:x '/x5C' x}{http://coq.inria.fr/distrib/8.5pl3/stdlib/Coq.Init.Logic}{\coqdocnotation{\ensuremath{\land}}} \coqdoceol
\coqdocindent{0.50em}
\coqexternalref{:type scope:x '/x5C' x}{http://coq.inria.fr/distrib/8.5pl3/stdlib/Coq.Init.Logic}{\coqdocnotation{(}}\coqdocvariable{r} \coqdocnotation{|>=}\coqdocvariable{t} \coqref{evidential.make}{\coqdocaxiom{make}} (\coqref{evidential.demand}{\coqdocdefinition{demand}} \coqdocvariable{r} \coqdocvariable{c} (\coqdocprojection{form} \coqdocvariable{d}) (\coqdocprojection{picture} \coqdocvariable{d}) (\coqdocprojection{pass} \coqdocvariable{d}) (\coqdocprojection{travels} \coqdocvariable{d})\coqdoceol
\coqdocindent{0.50em}
(\coqdocprojection{insurance} \coqdocvariable{d}) (\coqdocprojection{lodgings} \coqdocvariable{d}) (\coqdocprojection{sufficient} \coqdocvariable{d}) (\coqdocprojection{time\_stamp} \coqdocvariable{d}))\coqexternalref{:type scope:x '/x5C' x}{http://coq.inria.fr/distrib/8.5pl3/stdlib/Coq.Init.Logic}{\coqdocnotation{)}}.\coqdoceol
\coqdocemptyline
\end{coqdoccode}
The delivery policy states that if there is a valid demand which 
  corresponds to the visa made by the requester, the
  consulate can deliver the visa. The details of the validity of a demand
  are in appendix (see~\ref{sec:app}). It states that the seven requirements are
  satisfied and claimed by the appropriate authority. For example, the validity of 
  the passport of a citizen of a country $C$, is a claim of $C$, i.e. $C$ is accountable for 
  the validity of the passport. \begin{coqdoccode}
\coqdocemptyline
\coqdocnoindent
\coqdockw{Definition} \coqdef{evidential.delivering validation}{delivering\_validation}{\coqdocdefinition{delivering\_validation}} (\coqdocvar{cons} \coqdocvar{req}:\coqdocaxiom{Authority}) (\coqdocvar{v}:\coqdocaxiom{visa}) (\coqdocvar{t}:\coqdocdefinition{time}) := \coqdoceol
\coqdocindent{0.50em}
\coqexternalref{:type scope:'exists' x '..' x ',' x}{http://coq.inria.fr/distrib/8.5pl3/stdlib/Coq.Init.Logic}{\coqdocnotation{\ensuremath{\exists}}} \coqdocvar{d}\coqexternalref{:type scope:'exists' x '..' x ',' x}{http://coq.inria.fr/distrib/8.5pl3/stdlib/Coq.Init.Logic}{\coqdocnotation{,}} \coqref{evidential.visa of demand}{\coqdocaxiom{visa\_of\_demand}} \coqdocvariable{d} \coqdocvariable{v} \coqexternalref{:type scope:x '/x5C' x}{http://coq.inria.fr/distrib/8.5pl3/stdlib/Coq.Init.Logic}{\coqdocnotation{\ensuremath{\land}}} \coqref{evidential.demanding}{\coqdocdefinition{demanding}} \coqdocvariable{req} (\coqdocprojection{time\_stamp} \coqdocvariable{d}) (\coqdocaxiom{country\_of} \coqdocvariable{cons}) \coqdocvariable{d} \coqexternalref{:type scope:x '/x5C' x}{http://coq.inria.fr/distrib/8.5pl3/stdlib/Coq.Init.Logic}{\coqdocnotation{\ensuremath{\land}}} \coqdoceol
\coqdocindent{1.50em}
\coqdocdefinition{schengen\_demand\_validation} \coqdocvariable{req} \coqdocvariable{d} \coqexternalref{:type scope:x '/x5C' x}{http://coq.inria.fr/distrib/8.5pl3/stdlib/Coq.Init.Logic}{\coqdocnotation{\ensuremath{\land}}} \coqexternalref{:type scope:x '/x5C' x}{http://coq.inria.fr/distrib/8.5pl3/stdlib/Coq.Init.Logic}{\coqdocnotation{(}}\coqdocvariable{cons} \coqdocnotation{|>=}\coqdocvariable{t} \coqdocnotation{(}\coqref{evidential.make}{\coqdocaxiom{make}} (\coqref{evidential.deliver}{\coqdocdefinition{deliver}} \coqdocvariable{cons} \coqdocvariable{v})\coqdocnotation{)}\coqexternalref{:type scope:x '/x5C' x}{http://coq.inria.fr/distrib/8.5pl3/stdlib/Coq.Init.Logic}{\coqdocnotation{)}}.\coqdoceol
\coqdocemptyline
\coqdocemptyline
\coqdocnoindent
\coqdockw{Definition} \coqdef{evidential.delivering}{delivering}{\coqdocdefinition{delivering}} (\coqdocvar{cons} \coqdocvar{req}:\coqdocaxiom{Authority}) (\coqdocvar{v}:\coqdocaxiom{visa}) (\coqdocvar{t}:\coqdocdefinition{time}) := \coqdoceol
\coqdocindent{0.50em}
(\coqdocdefinition{D1} \coqdocvariable{req} \coqdocvariable{cons} (\coqdocvariable{cons} \coqdocnotation{|><}\coqdocvariable{t} \coqref{evidential.delivering validation}{\coqdocdefinition{delivering\_validation}} \coqdocvariable{cons} \coqdocvariable{req} \coqdocvariable{v} \coqdocvariable{t}) ).\coqdoceol
\coqdocemptyline
\end{coqdoccode}
Of course only specific persons can control a visa, this officer has to be one of
the Schengen area.
 \begin{coqdoccode} \coqdocemptyline \coqdocnoindent
  \coqdockw{Definition}
  \coqdef{evidential.controlling}{controlling}{\coqdocdefinition{controlling}}
  (\coqdocvar{officer}:\coqdocaxiom{Authority})
  (\coqdocvar{v}:\coqdocaxiom{visa})
  (\coqdocvar{t}:\coqdocdefinition{time}):= \coqdoceol
  \coqdocindent{1.50em} \coqref{evidential.schengen
    officer}{\coqdocaxiom{schengen\_officer}} \coqdocvariable{officer}
  \coqexternalref{:type scope:x '/x5C'
    x}{http://coq.inria.fr/distrib/8.5pl3/stdlib/Coq.Init.Logic}{\coqdocnotation{\ensuremath{\land}}}
  \coqdocdefinition{curr} \coqdocvariable{t} \coqexternalref{:type
    scope:x '/x5C'
    x}{http://coq.inria.fr/distrib/8.5pl3/stdlib/Coq.Init.Logic}{\coqdocnotation{\ensuremath{\land}}}
  \coqexternalref{:type scope:x '/x5C'
    x}{http://coq.inria.fr/distrib/8.5pl3/stdlib/Coq.Init.Logic}{\coqdocnotation{(}}\coqdocvariable{officer}
  \coqdocnotation{|>=}\coqdocvariable{t}
  \coqdocnotation{(}\coqref{evidential.make}{\coqdocaxiom{make}}
  (\coqref{evidential.control}{\coqdocdefinition{control}}
  \coqdocvariable{officer}
  \coqdocvariable{v})\coqdocnotation{)}\coqexternalref{:type scope:x
    '/x5C'
    x}{http://coq.inria.fr/distrib/8.5pl3/stdlib/Coq.Init.Logic}{\coqdocnotation{)}}.\coqdoceol
  \coqdocemptyline
\end{coqdoccode}
The suspicion policy states that an officer has some suspicions regarding 
  specific properties of the visa.

 First case, this was a false alert and the officer claims
  a proof that all queries are satisfied. \begin{coqdoccode}
\coqdocemptyline
\coqdocnoindent
\coqdockw{Definition} \coqdef{evidential.false alert}{false\_alert}{\coqdocdefinition{false\_alert}} (\coqdocvar{officer}:\coqdocaxiom{Authority}) (\coqdocvar{v}:\coqdocaxiom{visa})(\coqdocvar{ev}:\coqref{evidential.queries}{\coqdocdefinition{queries}}) (\coqdocvar{t}:\coqdocdefinition{time}):= \coqdoceol
\coqdocindent{0.50em}
\coqref{evidential.schengen officer}{\coqdocaxiom{schengen\_officer}} \coqdocvariable{officer} \coqexternalref{:type scope:x '/x5C' x}{http://coq.inria.fr/distrib/8.5pl3/stdlib/Coq.Init.Logic}{\coqdocnotation{\ensuremath{\land}}} \coqexternalref{:type scope:x '/x5C' x}{http://coq.inria.fr/distrib/8.5pl3/stdlib/Coq.Init.Logic}{\coqdocnotation{(}}\coqdocvariable{officer} \coqdocnotation{|>=}\coqdocvariable{t} \coqdocnotation{(}\coqref{evidential.make answer}{\coqdocaxiom{make\_answer}} (\coqref{evidential.suspect}{\coqdocdefinition{suspect}} \coqdocvariable{officer} \coqdocvariable{v} \coqdocvariable{ev})\coqdocnotation{)}\coqexternalref{:type scope:x '/x5C' x}{http://coq.inria.fr/distrib/8.5pl3/stdlib/Coq.Init.Logic}{\coqdocnotation{)}} \coqexternalref{:type scope:x '/x5C' x}{http://coq.inria.fr/distrib/8.5pl3/stdlib/Coq.Init.Logic}{\coqdocnotation{\ensuremath{\land}}}\coqdoceol
\coqdocindent{0.50em}
\coqref{evidential.suspect}{\coqdocdefinition{suspect}} \coqdocvariable{officer} \coqdocvariable{v} \coqdocvariable{ev}\coqexternalref{:type scope:x '=' x}{http://coq.inria.fr/distrib/8.5pl3/stdlib/Coq.Init.Logic}{\coqdocnotation{=}} \coqref{evidential.Valid}{\coqdocconstructor{Valid}}.\coqdoceol
\coqdocemptyline
\end{coqdoccode}
Second case, this was a real alert and the officer claims a list
  of suspicious claims that have to be checked. \begin{coqdoccode}
\coqdocemptyline
\coqdocnoindent
\coqdockw{Definition} \coqdef{evidential.raise alert}{raise\_alert}{\coqdocdefinition{raise\_alert}} (\coqdocvar{officer}:\coqdocaxiom{Authority})(\coqdocvar{v}:\coqdocaxiom{visa})(\coqdocvar{ev}:\coqref{evidential.queries}{\coqdocdefinition{queries}})
(\coqdocvar{evidences}: \coqexternalref{list}{http://coq.inria.fr/distrib/8.5pl3/stdlib/Coq.Init.Datatypes}{\coqdocaxiom{list}} \coqdockw{Prop}) (\coqdocvar{t}:\coqdocdefinition{time}):=\coqdoceol
\coqdocindent{0.50em}
\coqref{evidential.schengen officer}{\coqdocaxiom{schengen\_officer}} \coqdocvariable{officer} \coqexternalref{:type scope:x '/x5C' x}{http://coq.inria.fr/distrib/8.5pl3/stdlib/Coq.Init.Logic}{\coqdocnotation{\ensuremath{\land}}} \coqexternalref{:type scope:x '/x5C' x}{http://coq.inria.fr/distrib/8.5pl3/stdlib/Coq.Init.Logic}{\coqdocnotation{(}}\coqdocvariable{officer} \coqdocnotation{|>=}\coqdocvariable{t} \coqdocnotation{(}\coqref{evidential.make answer}{\coqdocaxiom{make\_answer}} (\coqref{evidential.suspect}{\coqdocdefinition{suspect}} \coqdocvariable{officer} \coqdocvariable{v} \coqdocvariable{ev})\coqdocnotation{)}\coqexternalref{:type scope:x '/x5C' x}{http://coq.inria.fr/distrib/8.5pl3/stdlib/Coq.Init.Logic}{\coqdocnotation{)}} \coqexternalref{:type scope:x '/x5C' x}{http://coq.inria.fr/distrib/8.5pl3/stdlib/Coq.Init.Logic}{\coqdocnotation{\ensuremath{\land}}} \coqdoceol
\coqdocindent{0.50em}
\coqref{evidential.suspect}{\coqdocdefinition{suspect}} \coqdocvariable{officer} \coqdocvariable{v} \coqdocvariable{ev}\coqexternalref{:type scope:x '=' x}{http://coq.inria.fr/distrib/8.5pl3/stdlib/Coq.Init.Logic}{\coqdocnotation{=}} \coqref{evidential.Suspects}{\coqdocconstructor{Suspects}} \coqdocvariable{evidences}.\coqdoceol
\end{coqdoccode}

\subsection{The Cyberlogic Protocols at Work}

In this section, we present a scenario specified in the Cyberlogic
theorem prover to highlight the accountability computation, its role
in conflict detection and the fact that it allows to raise an alert.
Today, this facility can only be applied at design time while playing
scenarios to improve Cyberlogic protocol. We plan to transpose this
facility at runtime via a monitoring smart contract or service. To do
this, the Cyberlogic framework has to be extended to consider
transactions as first class citizens and to express accountability on
transaction instead than only on properties. The scenario is as the
following:

  Jon Snow requests a Schengen visa to the French consulate for 3
months. He plans to stay in Paris from the 1st June 2018 to 31st
August 2018 in the Icy Wall.
\begin{coqdoccode}
\coqdocemptyline
\coqdocnoindent
\coqdockw{Definition} \coqdef{scene.JSacc}{JSacc}{\coqdocdefinition{JSacc}} := (\coqdocconstructor{mkAcc} \coqref{scene.IcyWall}{\coqdocaxiom{IcyWall}} \coqref{scene.FirstJune2018}{\coqdocaxiom{FirstJune2018}} \coqref{scene.ThirtyFirstAugust2018}{\coqdocaxiom{ThirtyFirstAugust2018}}).\coqdoceol
\coqdocnoindent
\coqdockw{Definition} \coqdef{scene.JSaccs}{JSaccs}{\coqdocdefinition{JSaccs}} := \coqref{scene.JSacc}{\coqdocdefinition{JSacc}}\coqexternalref{:list scope:x '::' x}{http://coq.inria.fr/distrib/8.5pl3/stdlib/Coq.Init.Datatypes}{\coqdocnotation{::}}\coqexternalref{nil}{http://coq.inria.fr/distrib/8.5pl3/stdlib/Coq.Init.Datatypes}{\coqdocconstructor{nil}}.\coqdoceol
\coqdocemptyline
\end{coqdoccode}
His flight tickets are:
  
  \begin{tabular}{cccc}
      Drogo airline & from Winterfell (Essos) & to Paris (France)& on
      the flight 3 \\
      & 1st June 2018 & departure : 3am & arrival 3:30am \\
      Drogo airline & from Paris (France) & to Winterfell
      (Essos) & on flight 10\\
      & 31st August 2018 & departure : 4pm & arrival 4:30pm \\
  \end{tabular}
  
\begin{coqdoccode}
\coqdocemptyline
\coqdocnoindent
\coqdockw{Definition} \coqdef{scene.JSoutward}{JSoutward}{\coqdocdefinition{JSoutward}} := \coqdoceol
\coqdocindent{0.50em}
(\coqdocconstructor{mkFlight} \coqref{scene.Drogo}{\coqdocaxiom{Drogo}} 3 \coqref{scene.JonSnow}{\coqdocaxiom{JonSnow}} \coqexternalref{:core scope:'(' x ',' x ',' '..' ',' x ')'}{http://coq.inria.fr/distrib/8.5pl3/stdlib/Coq.Init.Datatypes}{\coqdocnotation{(}}\coqref{scene.Winterfell}{\coqdocaxiom{Winterfell}}\coqexternalref{:core scope:'(' x ',' x ',' '..' ',' x ')'}{http://coq.inria.fr/distrib/8.5pl3/stdlib/Coq.Init.Datatypes}{\coqdocnotation{,}}\coqref{scene.FirstJune2018}{\coqdocaxiom{FirstJune2018}}\coqexternalref{:nat scope:x '+' x}{http://coq.inria.fr/distrib/8.5pl3/stdlib/Coq.Init.Peano}{\coqdocnotation{+}}\coqref{scene.Wdep t}{\coqdocaxiom{Wdep\_t}}\coqexternalref{:core scope:'(' x ',' x ',' '..' ',' x ')'}{http://coq.inria.fr/distrib/8.5pl3/stdlib/Coq.Init.Datatypes}{\coqdocnotation{)}} \coqdoceol
\coqdocindent{5.00em}
\coqexternalref{:core scope:'(' x ',' x ',' '..' ',' x ')'}{http://coq.inria.fr/distrib/8.5pl3/stdlib/Coq.Init.Datatypes}{\coqdocnotation{(}}\coqref{scene.France}{\coqdocaxiom{France}}\coqexternalref{:core scope:'(' x ',' x ',' '..' ',' x ')'}{http://coq.inria.fr/distrib/8.5pl3/stdlib/Coq.Init.Datatypes}{\coqdocnotation{,}}\coqref{scene.FirstJune2018}{\coqdocaxiom{FirstJune2018}}\coqexternalref{:nat scope:x '+' x}{http://coq.inria.fr/distrib/8.5pl3/stdlib/Coq.Init.Peano}{\coqdocnotation{+}}\coqref{scene.Farr t}{\coqdocaxiom{Farr\_t}}\coqexternalref{:core scope:'(' x ',' x ',' '..' ',' x ')'}{http://coq.inria.fr/distrib/8.5pl3/stdlib/Coq.Init.Datatypes}{\coqdocnotation{)}} \coqref{scene.W IATA}{\coqdocaxiom{W\_IATA}} \coqref{scene.F IATA}{\coqdocaxiom{F\_IATA}} 100).\coqdoceol
\coqdocnoindent
\coqdockw{Definition} \coqdef{scene.JSreturn}{JSreturn}{\coqdocdefinition{JSreturn}} := \coqdoceol
\coqdocindent{0.50em}
(\coqdocconstructor{mkFlight} \coqref{scene.Drogo}{\coqdocaxiom{Drogo}} 10 \coqref{scene.JonSnow}{\coqdocaxiom{JonSnow}} \coqexternalref{:core scope:'(' x ',' x ',' '..' ',' x ')'}{http://coq.inria.fr/distrib/8.5pl3/stdlib/Coq.Init.Datatypes}{\coqdocnotation{(}}\coqref{scene.France}{\coqdocaxiom{France}}\coqexternalref{:core scope:'(' x ',' x ',' '..' ',' x ')'}{http://coq.inria.fr/distrib/8.5pl3/stdlib/Coq.Init.Datatypes}{\coqdocnotation{,}}\coqref{scene.ThirtyFirstAugust2018}{\coqdocaxiom{ThirtyFirstAugust2018}}\coqexternalref{:nat scope:x '+' x}{http://coq.inria.fr/distrib/8.5pl3/stdlib/Coq.Init.Peano}{\coqdocnotation{+}}\coqref{scene.Fdep t}{\coqdocaxiom{Fdep\_t}}\coqexternalref{:core scope:'(' x ',' x ',' '..' ',' x ')'}{http://coq.inria.fr/distrib/8.5pl3/stdlib/Coq.Init.Datatypes}{\coqdocnotation{)}} \coqdoceol
\coqdocindent{5.50em}
\coqexternalref{:core scope:'(' x ',' x ',' '..' ',' x ')'}{http://coq.inria.fr/distrib/8.5pl3/stdlib/Coq.Init.Datatypes}{\coqdocnotation{(}}\coqref{scene.Winterfell}{\coqdocaxiom{Winterfell}}\coqexternalref{:core scope:'(' x ',' x ',' '..' ',' x ')'}{http://coq.inria.fr/distrib/8.5pl3/stdlib/Coq.Init.Datatypes}{\coqdocnotation{,}} \coqref{scene.ThirtyFirstAugust2018}{\coqdocaxiom{ThirtyFirstAugust2018}}\coqexternalref{:nat scope:x '+' x}{http://coq.inria.fr/distrib/8.5pl3/stdlib/Coq.Init.Peano}{\coqdocnotation{+}}\coqref{scene.Warr t}{\coqdocaxiom{Warr\_t}}\coqexternalref{:core scope:'(' x ',' x ',' '..' ',' x ')'}{http://coq.inria.fr/distrib/8.5pl3/stdlib/Coq.Init.Datatypes}{\coqdocnotation{)}} \coqref{scene.F IATA}{\coqdocaxiom{F\_IATA}} \coqref{scene.W IATA}{\coqdocaxiom{W\_IATA}} 100).\coqdoceol
\coqdocnoindent
\coqdockw{Definition} \coqdef{scene.JStravels}{JStravels}{\coqdocdefinition{JStravels}} := \coqref{scene.JSoutward}{\coqdocdefinition{JSoutward}}\coqexternalref{:list scope:x '::' x}{http://coq.inria.fr/distrib/8.5pl3/stdlib/Coq.Init.Datatypes}{\coqdocnotation{::}}\coqref{scene.JSreturn}{\coqdocdefinition{JSreturn}}\coqexternalref{:list scope:x '::' x}{http://coq.inria.fr/distrib/8.5pl3/stdlib/Coq.Init.Datatypes}{\coqdocnotation{::}}\coqexternalref{nil}{http://coq.inria.fr/distrib/8.5pl3/stdlib/Coq.Init.Datatypes}{\coqdocconstructor{nil}}.\coqdoceol
\coqdocemptyline
\end{coqdoccode}
   Thanks to his new job as King of The North, he provides his
   employment contract as a mean of sufficient.  
 \begin{coqdoccode}
\coqdocemptyline
\coqdocnoindent
\coqdockw{Definition} \coqdef{scene.was KingOfTheNorth}{was\_KingOfTheNorth}{\coqdocdefinition{was\_KingOfTheNorth}}:= \coqref{scene.KingOfTheNorth}{\coqdocaxiom{KingOfTheNorth}} \coqref{scene.JonSnow}{\coqdocaxiom{JonSnow}} \coqref{scene.demand t}{\coqdocaxiom{demand\_t}}.\coqdoceol
\coqdocnoindent
\coqdockw{Definition} \coqdef{scene.JSsuff}{JSsuff}{\coqdocdefinition{JSsuff}}:=  \coqdocconstructor{Employment} \coqref{scene.Cwinterfell}{\coqdocaxiom{Cwinterfell}} \coqref{scene.was KingOfTheNorth}{\coqdocdefinition{was\_KingOfTheNorth}}.\coqdoceol
\coqdocemptyline
\end{coqdoccode}
He also provides its passport number (delivered by Winterfell) and a
reference of its travel health insurance delivered by Three-eyed crow \&cie.). We have axiomatized the no significant part of the
requirements. 

\begin{coqdoccode} \coqdocemptyline \coqdocnoindent
  \coqdockw{Definition}
  \coqdef{scene.JSdemand}{JSdemand}{\coqdocdefinition{JSdemand}}:=
  (\coqdocconstructor{mkDemand}
  \coqref{scene.JSform}{\coqdocaxiom{JSform}}
  \coqref{scene.JSpic}{\coqdocaxiom{JSpic}}
  \coqref{scene.JSpassport}{\coqdocaxiom{JSpassport}}
  \coqref{scene.JStravels}{\coqdocdefinition{JStravels}}
  \coqref{scene.JSinsurance}{\coqdocaxiom{JSinsurance}}
  \coqref{scene.JSaccs}{\coqdocdefinition{JSaccs}}
  \coqref{scene.JSsuff}{\coqdocdefinition{JSsuff}}
  \coqref{scene.demand t}{\coqdocaxiom{demand\_t}}).\coqdoceol
  \coqdocemptyline
\end{coqdoccode}
He obtains his visa.\begin{coqdoccode}
\coqdocemptyline
\coqdocnoindent
\coqdockw{Variable} \coqdef{scene.JSvisa}{JSvisa}{\coqdocvariable{JSvisa}}: \coqdocaxiom{visa}.\coqdoceol
\coqdocemptyline
\end{coqdoccode}

    A season's later, the 1st July 2018, a police officer controls his
    visa.  This officer recognizes him and knows, thanks to the
    Winterfell Times that Lady Sansa Stark. It reveales that she took
    his job since 4 months. So he suspects his visas and asks for the
    mean of sufficient.
\begin{coqdoccode}
\coqdocemptyline
\coqdocnoindent
\coqdockw{Definition} \coqdef{scene.suspicious clue}{suspicious\_clue}{\coqdocdefinition{suspicious\_clue}} := \coqref{scene.WinterfellTime}{\coqdocaxiom{WinterfellTime}} \coqdocnotation{|>} \coqdocnotation{(}\coqref{scene.KingOfTheNorth}{\coqdocaxiom{KingOfTheNorth}} \coqref{scene.SansaStark}{\coqdocaxiom{SansaStark}} (\coqref{scene.demand t}{\coqdocaxiom{demand\_t}} \coqexternalref{:nat scope:x '-' x}{http://coq.inria.fr/distrib/8.5pl3/stdlib/Coq.Init.Peano}{\coqdocnotation{-}} 5)\coqdocnotation{)}.\coqdoceol
\coqdocemptyline
\coqdocnoindent
\coqdockw{Definition} \coqdef{scene.JSsuff query}{JSsuff\_query}{\coqdocdefinition{JSsuff\_query}} (\coqdocvar{v}:\coqdocaxiom{visa}):=\coqdoceol
\coqdocindent{0.50em}
\coqdockw{\ensuremath{\forall}} \coqdocvar{d}, \coqdocaxiom{visa\_of\_demand} \coqdocvariable{d} \coqdocvariable{v} \coqexternalref{:type scope:x '->' x}{http://coq.inria.fr/distrib/8.5pl3/stdlib/Coq.Init.Logic}{\coqdocnotation{\ensuremath{\rightarrow}}} \coqexternalref{:type scope:x '=' x}{http://coq.inria.fr/distrib/8.5pl3/stdlib/Coq.Init.Logic}{\coqdocnotation{(}}\coqdocprojection{sufficient} \coqdocvariable{d}\coqexternalref{:type scope:x '=' x}{http://coq.inria.fr/distrib/8.5pl3/stdlib/Coq.Init.Logic}{\coqdocnotation{)}} \coqexternalref{:type scope:x '=' x}{http://coq.inria.fr/distrib/8.5pl3/stdlib/Coq.Init.Logic}{\coqdocnotation{=}} \coqref{scene.JSsuff}{\coqdocdefinition{JSsuff}} \coqexternalref{:type scope:x '->' x}{http://coq.inria.fr/distrib/8.5pl3/stdlib/Coq.Init.Logic}{\coqdocnotation{\ensuremath{\rightarrow}}} 
\coqref{scene.KingOfTheNorth}{\coqdocaxiom{KingOfTheNorth}} \coqref{scene.JonSnow}{\coqdocaxiom{JonSnow}} \coqref{scene.demand t}{\coqdocaxiom{demand\_t}}.\coqdoceol
\coqdocemptyline
\end{coqdoccode}

   The protocol returns the claim of the employment contract, which is a claim of the Winterfell kingdom.

   At this stage, the smart contract waits for physical world intervention and raises an alert.
\begin{coqdoccode}
\coqdocemptyline
\coqdocnoindent
\coqdockw{Axiom} \coqdef{scene.suspicious sufficient means}{suspicious\_sufficient\_means}{\coqdocaxiom{suspicious\_sufficient\_means}}:\coqdoceol
\coqdocnoindent
\coqdocdefinition{demanding} \coqref{scene.JonSnow}{\coqdocaxiom{JonSnow}} \coqref{scene.demand t}{\coqdocaxiom{demand\_t}} \coqref{scene.France}{\coqdocaxiom{France}} \coqref{scene.JSdemand}{\coqdocdefinition{JSdemand}} \coqexternalref{:type scope:x '->' x}{http://coq.inria.fr/distrib/8.5pl3/stdlib/Coq.Init.Logic}{\coqdocnotation{\ensuremath{\rightarrow}}}
\coqdocdefinition{delivering} \coqref{scene.CFrance}{\coqdocaxiom{CFrance}} \coqref{scene.JonSnow}{\coqdocaxiom{JonSnow}} \coqref{scene.JSvisa}{\coqdocaxiom{JSvisa}} \coqref{scene.deliver t}{\coqdocaxiom{deliver\_t}} \coqexternalref{:type scope:x '->' x}{http://coq.inria.fr/distrib/8.5pl3/stdlib/Coq.Init.Logic}{\coqdocnotation{\ensuremath{\rightarrow}}}\coqdoceol
\coqdocnoindent
\coqdocdefinition{raise\_alert} \coqref{scene.JaimeL}{\coqdocaxiom{JaimeL}} \coqref{scene.JSvisa}{\coqdocaxiom{JSvisa}} (\coqref{scene.JSsuff query}{\coqdocdefinition{JSsuff\_query}}\coqexternalref{:list scope:x '::' x}{http://coq.inria.fr/distrib/8.5pl3/stdlib/Coq.Init.Datatypes}{\coqdocnotation{::}}\coqexternalref{nil}{http://coq.inria.fr/distrib/8.5pl3/stdlib/Coq.Init.Datatypes}{\coqdocconstructor{nil}})
(\coqexternalref{:list scope:x '::' x}{http://coq.inria.fr/distrib/8.5pl3/stdlib/Coq.Init.Datatypes}{\coqdocnotation{(}}\coqref{scene.Cwinterfell}{\coqdocaxiom{Cwinterfell}} \coqdocnotation{|>} \coqref{scene.was KingOfTheNorth}{\coqdocdefinition{was\_KingOfTheNorth}}\coqexternalref{:list scope:x '::' x}{http://coq.inria.fr/distrib/8.5pl3/stdlib/Coq.Init.Datatypes}{\coqdocnotation{)::}}\coqexternalref{nil}{http://coq.inria.fr/distrib/8.5pl3/stdlib/Coq.Init.Datatypes}{\coqdocconstructor{nil}}) \coqref{scene.suspicious t}{\coqdocaxiom{suspicious\_t}}.\coqdoceol
\end{coqdoccode}

While the expressiveness of Cyberlogic allows to specify and reason on
smart legal contracts and execution scenarios, a lot has to be done to
provide mechanisms to monitor accountability at runtime in a
blockchain-based execution environment \footnote{In
  section~\ref{sec:futurs} this issue is discussed as
  work-in-progress.}.  Moreover, the authorities algebra allows to
claim properties while in a a smart legal contract language, this is
the object/data of a transaction that has to be claimed. The chosen
example is an evidential protocol for which the Cyberlogic is
particularly suitable.  It also highlights the trust and accountability
issues that DAO has to manage.  However, to cover the whole kind of
DAO the Cyberlogic has to be extended to handle contractual
relationship (see section~\ref{sec:futurs}). The scenario we had
unfolded highlight the detection of conflict and the active control. To
treat more complex scenario, with several demands, the write, read and
query operations have semantics concording with the distributed and
immutable characteristics of the ledger. Therefore, those operations
have to be first-class citizen of the smart legal contract language.
The next section the blockchain-based smart contract, i.e. the code
counterpart of the Cyberlogic prototocol is presented.

\subsection{A solidity smart contract code}

We implemented the Shengen Visa smart contract given in
Section~\ref{sec:alg} in the Solidity language (Algorithm
\ref{alg:SolidityContract}). Solidity is a contract-oriented,
high-level language whose syntax is similar to that of JavaScript and
it is designed to target the Ethereum blockchain
framework~\cite{Dannen:2017:IES:3103305}.  It is statically-typed,
supports inheritance, libraries and user-defined abstract data types.
Solidity contracts bundle data with the functions operating on that
data and have mechanisms for restricting direct access to some of the
contracts's components.

The Shengen Visa smart contract functions are implemented as
\textsf{demand()}, \textsf{deliver()}, \textsf{control()} and
\textsf{suspect()} fuctions respectively in Algorithm
\ref{alg:SolidityContract}.  It is assummed that the identity
(address) of the consulate and the officer are already known by the
contract, and they are used by the contract to restrict access to its
functions.  External calls to the contract functions bring a
\textsf{msg.sender} parameter that returns the address of the caller.
The contract can then use these information to verify the identities
when needed using the \textsf{modifier} construct of Solidity together
with the \textsf{msg.sender} parameter. Modifiers are used to amend
the semantics of functions in a declarative way. The modifier
\textsf{onlyConsulate()} is used for verifying that
\textsf{msg.sender} is a valid consulate (Algorithm
\ref{alg:SolidityContract} line 6) and the modifier
\textsf{onlyOfficer()} used for verifying that \textsf{msg.sender} is
an officer of schengen aera (Algorithm \ref{alg:SolidityContract} line
7).  This way, it is assured that the \textsf{deliver()} function can
only be called by the consulate (Algorithm \ref{alg:SolidityContract}
line 18) and the \textsf{control()} and \textsf{suspect()} function
can only be called by the officer (Algorithm
\ref{alg:SolidityContract} line 30 and 34).

\begin{algorithm}[h!]
\caption{The soldity implementation of the Shengen Visa smart
  contract. It is assumed that the identity (address) of the consulate
  and the officer are already known by the contract. Due to the space
  limitation the abstract data types Demand and Visa are not shown.}
\label{alg:SolidityContract}
\begin{algorithmic}[1]		
   \STATE \textbf{contract} SchengenVisa $\{$
   \STATE    
   \STATE \textbf{address} \textbf{public} consulate;
   \STATE \textbf{address} \textbf{public} officer;
   \STATE 
   \STATE \textbf{modifier} onlyConsulate() $\{$ require(msg.sender == consulate); $\_$; $\}$
   \STATE \textbf{modifier} onlyOfficer() $\{$ require(msg.sender == officer); $\_$; $\}$
   \STATE
   \STATE \textbf{mapping} (\textbf{address} $=>$ Demand) demands;
   \STATE \textbf{mapping} (\textbf{uint} $=>$ Visa) visas;
   \STATE ...
   \STATE
   \STATE \textbf{function} demand(\textbf{uint} sch$\_$form$\_$id, \textbf{uint} photo$\_$id, \textbf{string} passport$\_$id, \textbf{uint}[] travel$\_$ids, \textbf{uint} travel$\_$health, \textbf{uint}[] accommodation$\_$ids, \textbf{uint} sufficient$\_$means, \textbf{uint} time$\_$stamp) \textbf{public} $\{$
	 \INDTWO \textbf{address} visaDemander = \textbf{msg.sender};     
     \INDTWO demands[visaDemander] = Demand(sch$\_$form$\_$id, photo$\_$id, passport$\_$id, travel$\_$ids,
     \INDFOUR travel$\_$health, accommodation$\_$ids, sufficient$\_$means, time$\_$stamp); $\}$
   \STATE 
   \STATE \textbf{function} deliver(Demand demand, \textbf{string} country, string duration) \textbf{public} \textbf{returns} (\textbf{uint} visaId) onlyConsulate $\{$
     \INDTWO \textbf{if} (isValid(demand)) $\{$
     \INDFOUR visas[++visaId] = Visa(visaId, consulate, country, duration);
     \INDFOUR \textbf{return} visa.id;
     \INDTWO $\}$ \textbf{else return} -1; $\}$ 
   \STATE 
   \STATE \textbf{function} isValid(Demand demand) \textbf{returns} (\textbf{bool} valid) \textbf{private} $\{$
   	 \INDTWO valid = validateTravels(demand.travel$\_$ids, demand.accommodation$\_$ids);
   	 \INDTWO valid $\&$= (demand.sufficient$\_$means $>$= 5000);
   	 \INDTWO \textbf{return} valid;
   \STATE $\}$
   \STATE
   \STATE \textbf{function} control(\textbf{uint} visaId) \textbf{public} \textbf{returns} (Visa visa) onlyOfficer $\{$
      \INDTWO \textbf{var} visa = visas[visaId];
      \INDTWO \textbf{return} visa; $\}$
   \STATE 
   \STATE \textbf{function} suspect(\textbf{uint} visaId, \textbf{string} reqField) \textbf{public}  \textbf{returns} (\textbf{var} field) onlyOfficer $\{$
      \INDTWO \textbf{var} visa = visas[visaId];
      \INDTWO \textbf{var} field = visa[reqField];
      \INDTWO \textbf{return} field; $\}$
   \STATE
   \STATE $\}$ // \textit{contract ShengenVisa}
\end{algorithmic} 
\end{algorithm}

\section{On-going and Futures Research Perspectives}{\label{sec:futurs}}

 Our aim is to provide a framework to design and implement smart
 contracts in a secure and trusted manner. Our point of view is that a
 smart legal contract has to be compiled in a smart contract code. We
 advocate that an extension of Cyberlogic is a good candidate to specify 
 the smart legal contract. As transparency and trust are at core of our
 vision, we plan to equip our framework with formal analyzers as
 well as a formal verification of the compiler. In this section, we
 discuss the different current works in progress and futures research
perspectives that we intend to explore.

\subsection{Specification of Legal Artefacts}

Smart contracts aim at digitized parts of legal contracts between
entities. Specifying law is a recurrent holy Grail in formal
methods~\cite{Prakken2002}. However, deontic
logic~\cite{sep-logic-deontic} with some restriction to avoid
paradoxes have been embedded in formal frameworks. In particular, the
Contract Language~\cite{Fenech2009} is an action-based language
featuring concepts for permission, obligation and prohibition as
first-class citizen. The tool CLAN allows to analyze specification in
CL and detect conflicts. On the other hand, CL does not allow to reason about trust and accountability
as Cyberlogic does. We aim at extending Cyberlogic with CL-like mechanisms to provide more specific expressiveness
dedicated to smart legal contracts.  The open project
~\href{ttps://legalese.com}{Legalese} also aims at providing such a
language based on {\em Contract Language}.

\subsection{Targeting Secure Smart Contract Code}

Cyberlogic protocols in our theorem prover in Coq are Gallinea
programs. Therefore, they can be extracted to Ocaml code and then be
compiled into a smart contract code language. In particular, we target
Ocaml based languages as they might provide a formal base such as
Michelson for the smart contract language of
Tezos~\cite{tezos-white-paper}. Another possibility is to target a
subset of Solidity~\cite{Dannen:2017:IES:3103305} of the Ethereum
Virtual Machine EVM~\cite{wood2014ethereum}.  The execution
infrastructure should also include a monitoring mechanism for
trace/scenarios analysis which is linked to the Cyberlogic
formalisation and allowing reasoning.

\subsection{Specification of the Blockchain at Low-Level}

The inherent non-determinism caused by the execution of the programs
in a widely distributed environment subject to Byzantine failures and
network partitions exposes the programs to several vulnerabilities,
which are often misunderstood by programmers. The survey~\cite{Atzei1}
shows, indeed, that most attacks in Ethereum were caused by
vulnerabilities of the execution platform (Ethereum Virtual Machine
and blockchain protocols).

Available formal analysis of security of the
blockchain~\cite{garay2015bitcoin} have several limitations, assuming
a perfect message diffusion mechanism, instantaneous communication and
a fixed number of participants. These assumptions are far from being
realistic so security thresholds as the famous ``majority assumption"
for Bitcoin (the system is secure if the majority the hashing power is
in the hands of honest nodes) fails shortly in more realistic
settings. A formal definition of blockchain (first attempts
in~\cite{DBLP:conf/sss/AnceaumeLPT17, DBLP:journals/corr/CrainGLR17})
is indeed needed to (i) allow the secure design of blockchain
protocols and (ii) to gain trust in the smart contract execution by
defining formal semantics to specify the properties of the execution
context.

\section{Conclusion}

We advocate that since the final goal of smart contracts is to obviate
the use of trusted third-parties, a smart contract specification must
allow reasoning about trust and accountability. By considering smart
contract as evidential protocols we take advantage of Cyberlogic to
specify and verify them.  In this paper we have shown the use of
Cyberlogic through an illustrative example and we have described the
overall approach along with the elements needed to provide a
trustworthy framework to design and implement secure and
trusted-by-design smart contracts. Finally, we are currently exploring
extensions of Cyberlogic with deontic supports and dedicated features
to interact with the underling immutable distributed ledger.

\bibliographystyle{abbrv} \bibliography{biblio}

\begin{thebibliography}{10}

\bibitem{DBLP:conf/sss/AnceaumeLPT17}
E.~Anceaume, R.~Ludinard, M.~Potop{-}Butucaru, and F.~Tronel.
\newblock Bitcoin a distributed shared register.
\newblock In {\em Stabilization, Safety, and Security of Distributed Systems -
  19th International Symposium, {SSS} 2017, Boston, MA, USA, November 5-8,
  2017, Proceedings}, pages 456--468, 2017.

\bibitem{Appel1999}
A.~W. Appel and E.~W. Felten.
\newblock Proof-carrying authentication.
\newblock In {\em Proceedings of the 6th ACM Conference on Computer and
  Communications Security}, CCS '99, pages 52--62, New York, NY, USA, 1999.
  ACM.

\bibitem{Atzei1}
N.~Atzei, M.~Bartoletti, and T.~Cimoli.
\newblock A survey of attacks on ethereum smart contracts sok.
\newblock In {\em Conference on Principles of Security and Trust - Volume
  10204}, New York, USA, 2017.

\bibitem{CyberL}
V.~Bernat, H.~Ruess, and N.~Shankar.
\newblock First-order cyberlogic.
\newblock Technical report, SRI International, 2005.

\bibitem{Bhargavan:2016:FVS:2993600.2993611}
K.~Bhargavan, A.~Delignat-Lavaud, C.~Fournet, A.~Gollamudi, G.~Gonthier,
  N.~Kobeissi, N.~Kulatova, A.~Rastogi, T.~Sibut-Pinote, N.~Swamy, and
  S.~Zanella-B{\'e}guelin.
\newblock Formal verification of smart contracts.
\newblock In {\em Workshop on Programming Languages and Analysis for Security},
  PLAS '16, New York, USA, 2016.

\bibitem{DBLP:conf/spw/BlazeFK98}
M.~Blaze, J.~Feigenbaum, and A.~D. Keromytis.
\newblock Keynote: Trust management for public-key infrastructures (position
  paper).
\newblock In {\em Security Protocols, 6th International Workshop, Cambridge,
  UK, April 15-17, 1998, Proceedings}, pages 59--63, 1998.

\bibitem{Blaze98}
M.~Blaze, J.~Feigenbaum, and M.~Strauss.
\newblock Compliance checking in the policymaker trust management system.
\newblock pages 254--274. Springer, 1998.

\bibitem{DBLP:journals/corr/CrainGLR17}
T.~Crain, V.~Gramoli, M.~Larrea, and M.~Raynal.
\newblock (leader/randomization/signature)-free byzantine consensus for
  consortium blockchains.
\newblock {\em CoRR}, abs/1702.03068, 2017.

\bibitem{theDAO}
P.~Daian.
\newblock Analysis of the dao exploit.
\newblock
  \url{http://hackingdistributed.com/2016/06/18/analysis-of-the-dao-exploit/}.

\bibitem{Dannen:2017:IES:3103305}
C.~Dannen.
\newblock {\em Introducing Ethereum and Solidity: Foundations of Cryptocurrency
  and Blockchain Programming for Beginners}.
\newblock Apress, Berkely, CA, USA, 1st edition, 2017.

\bibitem{Fenech2009}
S.~Fenech, G.~J. Pace, and G.~Schneider.
\newblock {\em CLAN: A Tool for Contract Analysis and Conflict Discovery},
  pages 90--96.
\newblock Springer Berlin Heidelberg, Berlin, Heidelberg, 2009.

\bibitem{garay2015bitcoin}
J.~A. Garay, A.~Kiayias, and N.~Leonardos.
\newblock The bitcoin backbone protocol: Analysis and applications.
\newblock In {\em EUROCRYPT (2)}, pages 281--310, 2015.

\bibitem{tezos-white-paper}
L.~Goodman.
\newblock A self-amending crypto-ledger. tezos white paper.
\newblock 2014.

\bibitem{LGF00}
N.~Li, B.~N. Grosof, and J.~Feigenbaum.
\newblock A practically implementable and tractable {Delegation Logic}.
\newblock In {\em Proceedings of the 2000 IEEE Symposium on Security and
  Privacy}, pages 27--42. IEEE Computer Society Press, May 2000.

\bibitem{Coq:manual}
\mbox{The Coq development team}.
\newblock {\em The Coq proof assistant reference manual}.
\newblock LogiCal Project, 2004.
\newblock Ver. 8.0.

\bibitem{sep-logic-deontic}
P.~McNamara.
\newblock Deontic logic.
\newblock In E.~N. Zalta, editor, {\em The Stanford Encyclopedia of
  Philosophy}. Metaphysics Research Lab, Stanford University, winter 2014
  edition, 2014.

\bibitem{pvs}
S.~Owre, J.~M. Rushby, and N.~Shankar.
\newblock Pvs: A prototype verification system.
\newblock In {\em Conference on Automated Deduction: Automated Deduction},
  CADE-11, pages 748--752, London, UK, UK, 1992. Springer-Verlag.

\bibitem{Prakken2002}
H.~Prakken and G.~Sartor.
\newblock {\em The Role of Logic in Computational Models of Legal Argument: A
  Critical Survey}, pages 342--381.
\newblock Springer Berlin Heidelberg, Berlin, Heidelberg, 2002.

\bibitem{RS:HCSS03}
Rue{\ss} and N.~Shankar.
\newblock Introducing cyberlogic.
\newblock In B.~Martin, editor, {\em HCSS'03---High Confidence Software and
  Systems Conference}, Baltimore, MD, 1-3 April 2003.

\bibitem{JStark}
J.~Stark.
\newblock Making sense of blockchain smart contracts.
\newblock \url{https://www.coindesk.com/making-sense-smart-contracts/}.

\bibitem{SCdef}
D.~Tapscott and A.~Tapscott.
\newblock {\em The blockchain revolution:how the technology behind bitcoin is
  changing Money,Business and the World}, pages 72,88,101,127.
\newblock TNew York, New York : Portfolio / Penguin, 2016.
\newblock ISBN-13: 978-1101980132.

\bibitem{wood2014ethereum}
G.~Wood.
\newblock Ethereum: A secure decentralised generalised transaction ledger.
\newblock
  \url{http://bitcoinaffiliatelist.com/wp-content/uploads/ethereum.pdf}, 2014.
\newblock Accessed: 2016-08-22.

\bibitem{Fidelis}
W.~Yao.
\newblock {Trust management for widely distributed systems}.
\newblock Technical Report UCAM-CL-TR-608, University of Cambridge, Computer
  Laboratory, Nov. 2004.

\end{thebibliography}

\appendix
\section{Specification of the Verification of Schengen Visa Requirements}{\label{sec:app}}

 The Schengen visa is an evidential protocols. In this section, we
  implement it as a Cyberlogic protocols. For the sake of simplicity, we had
  specified most of the components axiomatically. 
\begin{coqdoccode}
\coqdocemptyline
\coqdocemptyline
\end{coqdoccode}
\subsection{Specification and Properties of Data} 

 A smart contract for Schengen visa requests a generic specification
  of visa. A visa is delivered by a specific authority, has a
  duration or an expiry date, is an evidence that allow someone to
  circulate in a country.

\begin{coqdoccode}
\coqdocemptyline
\coqdocnoindent
\coqdockw{Variable} \coqdef{sch.visa}{visa}{\coqdocvariable{visa}}: \coqdockw{Type}.\coqdoceol
\coqdocnoindent
\coqdockw{Variable} \coqdef{sch.visa delivered by}{visa\_delivered\_by}{\coqdocvariable{visa\_delivered\_by}} : \coqref{sch.visa}{\coqdocaxiom{visa}} \coqexternalref{:type scope:x '->' x}{http://coq.inria.fr/distrib/8.5pl3/stdlib/Coq.Init.Logic}{\coqdocnotation{\ensuremath{\rightarrow}}} \coqdocaxiom{cyber.Authority}.\coqdoceol
\coqdocnoindent
\coqdockw{Variable} \coqdef{sch.visa duration}{visa\_duration}{\coqdocvariable{visa\_duration}}: \coqref{sch.visa}{\coqdocaxiom{visa}} \coqexternalref{:type scope:x '->' x}{http://coq.inria.fr/distrib/8.5pl3/stdlib/Coq.Init.Logic}{\coqdocnotation{\ensuremath{\rightarrow}}} \coqdocdefinition{cyber.time}.\coqdoceol
\coqdocnoindent
\coqdockw{Variable} \coqdef{sch.visa kind}{visa\_kind}{\coqdocvariable{visa\_kind}} : \coqref{sch.visa}{\coqdocaxiom{visa}} \coqexternalref{:type scope:x '->' x}{http://coq.inria.fr/distrib/8.5pl3/stdlib/Coq.Init.Logic}{\coqdocnotation{\ensuremath{\rightarrow}}} \coqdockw{Prop}.\coqdoceol
\coqdocnoindent
\coqdockw{Variable} \coqdef{sch.visa country}{visa\_country}{\coqdocvariable{visa\_country}}: \coqref{sch.visa}{\coqdocaxiom{visa}} \coqexternalref{:type scope:x '->' x}{http://coq.inria.fr/distrib/8.5pl3/stdlib/Coq.Init.Logic}{\coqdocnotation{\ensuremath{\rightarrow}}} \coqref{sch.country}{\coqdocdefinition{country}}.\coqdoceol
\coqdocemptyline
\end{coqdoccode}
{\it 1-The Visa Application form must be fully completed and
signed in the corresponding blanks.} 
\begin{coqdoccode}
\coqdocemptyline
\coqdocnoindent
\coqdockw{Variable} \coqdef{sch.schengen form}{schengen\_form}{\coqdocvariable{schengen\_form}}:\coqdockw{Type}.\coqdoceol
\coqdocnoindent
\coqdockw{Variable} \coqdef{sch.schengen from}{schengen\_from}{\coqdocvariable{schengen\_from}}:\coqref{sch.schengen form}{\coqdocaxiom{schengen\_form}} \coqexternalref{:type scope:x '->' x}{http://coq.inria.fr/distrib/8.5pl3/stdlib/Coq.Init.Logic}{\coqdocnotation{\ensuremath{\rightarrow}}} \coqdocdefinition{cyber.time}.\coqdoceol
\coqdocnoindent
\coqdockw{Variable} \coqdef{sch.schengen to}{schengen\_to}{\coqdocvariable{schengen\_to}}: \coqref{sch.schengen form}{\coqdocaxiom{schengen\_form}} \coqexternalref{:type scope:x '->' x}{http://coq.inria.fr/distrib/8.5pl3/stdlib/Coq.Init.Logic}{\coqdocnotation{\ensuremath{\rightarrow}}} \coqdocdefinition{cyber.time}.\coqdoceol
\coqdocnoindent
\coqdockw{Variable} \coqdef{sch.schengen requester}{schengen\_requester}{\coqdocvariable{schengen\_requester}}:\coqref{sch.schengen form}{\coqdocaxiom{schengen\_form}} \coqexternalref{:type scope:x '->' x}{http://coq.inria.fr/distrib/8.5pl3/stdlib/Coq.Init.Logic}{\coqdocnotation{\ensuremath{\rightarrow}}} \coqdocaxiom{cyber.Authority}.\coqdoceol
\coqdocnoindent
\coqdockw{Variable} \coqdef{sch.schengen country}{schengen\_country}{\coqdocvariable{schengen\_country}} : \coqref{sch.schengen form}{\coqdocaxiom{schengen\_form}} \coqexternalref{:type scope:x '->' x}{http://coq.inria.fr/distrib/8.5pl3/stdlib/Coq.Init.Logic}{\coqdocnotation{\ensuremath{\rightarrow}}} \coqref{sch.country}{\coqdocdefinition{country}}.\coqdoceol
\coqdocemptyline
\end{coqdoccode}
schengen\_form\_requirement holds if the form is valid.\begin{coqdoccode}
\coqdocemptyline
\coqdocnoindent
\coqdockw{Variable} \coqdef{sch.schengen form requirement}{schengen\_form\_requirement}{\coqdocvariable{schengen\_form\_requirement}}: \coqref{sch.schengen form}{\coqdocaxiom{schengen\_form}} \coqexternalref{:type scope:x '->' x}{http://coq.inria.fr/distrib/8.5pl3/stdlib/Coq.Init.Logic}{\coqdocnotation{\ensuremath{\rightarrow}}} \coqdockw{Prop}.\coqdoceol
\coqdocemptyline
\coqdocemptyline
\coqdocnoindent
\coqdockw{Variable} \coqdef{sch.photo}{photo}{\coqdocvariable{photo}}: \coqdockw{Type}.\coqdoceol
\coqdocnoindent
\coqdockw{Variable} \coqdef{sch.passport photo}{passport\_photo}{\coqdocvariable{passport\_photo}}: \coqref{sch.photo}{\coqdocaxiom{photo}} \coqexternalref{:type scope:x '->' x}{http://coq.inria.fr/distrib/8.5pl3/stdlib/Coq.Init.Logic}{\coqdocnotation{\ensuremath{\rightarrow}}} \coqdockw{Prop}.\coqdoceol
\coqdocemptyline
\end{coqdoccode}
{\it 2- A passport is delivered by an authority, has a expiry date and
contains a list of visas.}
\begin{coqdoccode}
\coqdocemptyline
\coqdocnoindent
\coqdockw{Variable} \coqdef{sch.passport}{passport}{\coqdocvariable{passport}}:\coqdockw{Type}.\coqdoceol
\coqdocnoindent
\coqdockw{Variable} \coqdef{sch.passport delivered by}{passport\_delivered\_by}{\coqdocvariable{passport\_delivered\_by}}: \coqref{sch.passport}{\coqdocaxiom{passport}} \coqexternalref{:type scope:x '->' x}{http://coq.inria.fr/distrib/8.5pl3/stdlib/Coq.Init.Logic}{\coqdocnotation{\ensuremath{\rightarrow}}} \coqdocaxiom{cyber.Authority}.\coqdoceol
\coqdocnoindent
\coqdockw{Variable} \coqdef{sch.visas of passport}{visas\_of\_passport}{\coqdocvariable{visas\_of\_passport}} : \coqexternalref{list}{http://coq.inria.fr/distrib/8.5pl3/stdlib/Coq.Init.Datatypes}{\coqdocinductive{list}} \coqref{sch.visa}{\coqdocaxiom{visa}}.\coqdoceol
\coqdocnoindent
\coqdockw{Variable} \coqdef{sch.passport expericy date}{passport\_expericy\_date}{\coqdocvariable{passport\_expericy\_date}}: \coqref{sch.passport}{\coqdocaxiom{passport}} \coqexternalref{:type scope:x '->' x}{http://coq.inria.fr/distrib/8.5pl3/stdlib/Coq.Init.Logic}{\coqdocnotation{\ensuremath{\rightarrow}}} \coqdocdefinition{cyber.time}.\coqdoceol
\coqdocemptyline
\end{coqdoccode}
Passport\_of is a mapping that associates a passport to its
owner.\begin{coqdoccode}
\coqdocemptyline
\coqdocnoindent
\coqdockw{Variable} \coqdef{sch.passport of}{passport\_of}{\coqdocvariable{passport\_of}}: \coqdocaxiom{cyber.Authority} \coqexternalref{:type scope:x '->' x}{http://coq.inria.fr/distrib/8.5pl3/stdlib/Coq.Init.Logic}{\coqdocnotation{\ensuremath{\rightarrow}}} \coqref{sch.passport}{\coqdocaxiom{passport}}.\coqdoceol
\coqdocemptyline
\end{coqdoccode}
{\it 3- The passport as well as all the copies of your previous
    visas, valid for at least 3 months prior to your departure is
    required. The passport must have at least two blank pages.}
\begin{coqdoccode}
\coqdocemptyline
\coqdocnoindent
\coqdockw{Definition} \coqdef{sch.valid passport at}{valid\_passport\_at}{\coqdocdefinition{valid\_passport\_at}} (\coqdocvar{p}:\coqref{sch.passport}{\coqdocaxiom{passport}}) (\coqdocvar{departure\_time}: \coqdocdefinition{cyber.time}):=\coqdoceol
\coqdocindent{0.50em}
\coqexternalref{:nat scope:x '<=' x}{http://coq.inria.fr/distrib/8.5pl3/stdlib/Coq.Init.Peano}{\coqdocnotation{(}}\coqref{sch.passport expericy date}{\coqdocaxiom{passport\_expericy\_date}} \coqdocvariable{p}\coqexternalref{:nat scope:x '<=' x}{http://coq.inria.fr/distrib/8.5pl3/stdlib/Coq.Init.Peano}{\coqdocnotation{)}} \coqexternalref{:nat scope:x '<=' x}{http://coq.inria.fr/distrib/8.5pl3/stdlib/Coq.Init.Peano}{\coqdocnotation{\ensuremath{\le}}} \coqexternalref{:nat scope:x '<=' x}{http://coq.inria.fr/distrib/8.5pl3/stdlib/Coq.Init.Peano}{\coqdocnotation{(}}\coqdocvariable{departure\_time} \coqexternalref{:nat scope:x '-' x}{http://coq.inria.fr/distrib/8.5pl3/stdlib/Coq.Init.Peano}{\coqdocnotation{-}} \coqexternalref{:nat scope:x '-' x}{http://coq.inria.fr/distrib/8.5pl3/stdlib/Coq.Init.Peano}{\coqdocnotation{(}}\coqref{sch.months}{\coqdocaxiom{months}} 3\coqexternalref{:nat scope:x '-' x}{http://coq.inria.fr/distrib/8.5pl3/stdlib/Coq.Init.Peano}{\coqdocnotation{)}}\coqexternalref{:nat scope:x '<=' x}{http://coq.inria.fr/distrib/8.5pl3/stdlib/Coq.Init.Peano}{\coqdocnotation{)}}.\coqdoceol
\coqdocemptyline
\coqdocnoindent
\coqdockw{Variable} \coqdef{sch.valid passport}{valid\_passport}{\coqdocvariable{valid\_passport}}: \coqref{sch.passport}{\coqdocaxiom{passport}} \coqexternalref{:type scope:x '->' x}{http://coq.inria.fr/distrib/8.5pl3/stdlib/Coq.Init.Logic}{\coqdocnotation{\ensuremath{\rightarrow}}} \coqdocdefinition{cyber.time} \coqexternalref{:type scope:x '->' x}{http://coq.inria.fr/distrib/8.5pl3/stdlib/Coq.Init.Logic}{\coqdocnotation{\ensuremath{\rightarrow}}} \coqdockw{Prop}.\coqdoceol
\coqdocnoindent
\coqdockw{Definition} \coqdef{sch.valid at time}{valid\_at\_time}{\coqdocdefinition{valid\_at\_time}} (\coqdocvar{p}:\coqref{sch.passport}{\coqdocaxiom{passport}})(\coqdocvar{departure\_time}: \coqdocdefinition{cyber.time}):=\coqdoceol
\coqdocindent{0.50em}
\coqexternalref{:type scope:x '->' x}{http://coq.inria.fr/distrib/8.5pl3/stdlib/Coq.Init.Logic}{\coqdocnotation{(}}\coqref{sch.valid passport}{\coqdocaxiom{valid\_passport}} \coqdocvariable{p} \coqdocvariable{departure\_time}\coqexternalref{:type scope:x '->' x}{http://coq.inria.fr/distrib/8.5pl3/stdlib/Coq.Init.Logic}{\coqdocnotation{)}} \coqexternalref{:type scope:x '->' x}{http://coq.inria.fr/distrib/8.5pl3/stdlib/Coq.Init.Logic}{\coqdocnotation{\ensuremath{\rightarrow}}} \coqexternalref{:type scope:x '->' x}{http://coq.inria.fr/distrib/8.5pl3/stdlib/Coq.Init.Logic}{\coqdocnotation{(}}\coqref{sch.valid passport at}{\coqdocdefinition{valid\_passport\_at}} \coqdocvariable{p} \coqdocvariable{departure\_time}\coqexternalref{:type scope:x '->' x}{http://coq.inria.fr/distrib/8.5pl3/stdlib/Coq.Init.Logic}{\coqdocnotation{)}}.\coqdoceol
\coqdocemptyline
\end{coqdoccode}
{\it 4- Round trip reservation or itinerary with dates and flight
  numbers specifying entry and exit from the Schengen area. You can
  use the visa consultation services like this one. These guys can
  handle most of your visa requirements such as flight itineraries,
  hotel reservations along with free consultation over email.}

\begin{coqdoccode}
\coqdocemptyline
\coqdocnoindent
\coqdockw{Definition} \coqdef{sch.IATA}{IATA}{\coqdocdefinition{IATA}}:=\coqexternalref{nat}{http://coq.inria.fr/distrib/8.5pl3/stdlib/Coq.Init.Datatypes}{\coqdocinductive{nat}}.\coqdoceol
\coqdocemptyline
\coqdocnoindent
\coqdockw{Record} \coqdef{sch.flight}{flight}{\coqdocrecord{flight}} := \coqdef{sch.mkFlight}{mkFlight}{\coqdocconstructor{mkFlight}}\coqdoceol
\coqdocnoindent
\{ \coqdef{sch.fl airline}{fl\_airline}{\coqdocprojection{fl\_airline}}: \coqdocaxiom{cyber.Authority}; \coqdoceol
\coqdocindent{1.00em}
\coqdef{sch.fl id}{fl\_id}{\coqdocprojection{fl\_id}}: \coqexternalref{nat}{http://coq.inria.fr/distrib/8.5pl3/stdlib/Coq.Init.Datatypes}{\coqdocinductive{nat}}; \coqdoceol
\coqdocindent{1.00em}
\coqdef{sch.fl for}{fl\_for}{\coqdocprojection{fl\_for}}: \coqdocaxiom{cyber.Authority}; \coqdoceol
\coqdocindent{1.00em}
\coqdef{sch.fl departure}{fl\_departure}{\coqdocprojection{fl\_departure}}: \coqref{sch.country}{\coqdocdefinition{country}} \coqexternalref{:type scope:x '*' x}{http://coq.inria.fr/distrib/8.5pl3/stdlib/Coq.Init.Datatypes}{\coqdocnotation{\ensuremath{\times}}} \coqdocdefinition{cyber.time}; \coqdoceol
\coqdocindent{1.00em}
\coqdef{sch.fl arrival}{fl\_arrival}{\coqdocprojection{fl\_arrival}}: \coqref{sch.country}{\coqdocdefinition{country}} \coqexternalref{:type scope:x '*' x}{http://coq.inria.fr/distrib/8.5pl3/stdlib/Coq.Init.Datatypes}{\coqdocnotation{\ensuremath{\times}}} \coqdocdefinition{cyber.time}; \coqdoceol
\coqdocindent{1.00em}
\coqdef{sch.fl dep aipport}{fl\_dep\_aipport}{\coqdocprojection{fl\_dep\_aipport}}: \coqref{sch.IATA}{\coqdocdefinition{IATA}}; \coqdoceol
\coqdocindent{1.00em}
\coqdef{sch.fl arr airport}{fl\_arr\_airport}{\coqdocprojection{fl\_arr\_airport}}:\coqref{sch.IATA}{\coqdocdefinition{IATA}};\coqdoceol
\coqdocindent{1.00em}
\coqdef{sch.fl price}{fl\_price}{\coqdocprojection{fl\_price}}: \coqexternalref{nat}{http://coq.inria.fr/distrib/8.5pl3/stdlib/Coq.Init.Datatypes}{\coqdocinductive{nat}}; \coqdoceol
\coqdocindent{0.50em}
\}.\coqdoceol
\coqdocemptyline
\coqdocnoindent
\coqdockw{Definition} \coqdef{sch.travel itinerary}{travel\_itinerary}{\coqdocdefinition{travel\_itinerary}} := \coqexternalref{list}{http://coq.inria.fr/distrib/8.5pl3/stdlib/Coq.Init.Datatypes}{\coqdocinductive{list}} \coqref{sch.flight}{\coqdocrecord{flight}}.\coqdoceol
\coqdocemptyline
\end{coqdoccode}
travel\_valid holds is a flight that is valid recording the
  requirement of the Schengen visa. That would be a property
  claimed by an authority. \begin{coqdoccode}
\coqdocemptyline
\coqdocnoindent
\coqdockw{Variable} \coqdef{sch.travel valid}{travel\_valid}{\coqdocvariable{travel\_valid}}: \coqref{sch.flight}{\coqdocrecord{flight}} \coqexternalref{:type scope:x '->' x}{http://coq.inria.fr/distrib/8.5pl3/stdlib/Coq.Init.Logic}{\coqdocnotation{\ensuremath{\rightarrow}}} \coqdockw{Prop}.\coqdoceol
\coqdocnoindent
\coqdockw{Definition} \coqdef{sch.travels validation}{travels\_validation}{\coqdocdefinition{travels\_validation}} (\coqdocvar{l}:\coqref{sch.travel itinerary}{\coqdocdefinition{travel\_itinerary}}) := \coqdoceol
\coqdocindent{0.50em}
\coqdockw{\ensuremath{\forall}} \coqdocvar{fl}, \coqexternalref{In}{http://coq.inria.fr/distrib/8.5pl3/stdlib/Coq.Lists.List}{\coqdocdefinition{List.In}} \coqdocvariable{fl} \coqdocvariable{l} \coqexternalref{:type scope:x '->' x}{http://coq.inria.fr/distrib/8.5pl3/stdlib/Coq.Init.Logic}{\coqdocnotation{\ensuremath{\rightarrow}}} \coqexternalref{:type scope:x '->' x}{http://coq.inria.fr/distrib/8.5pl3/stdlib/Coq.Init.Logic}{\coqdocnotation{(}}\coqdocnotation{(}\coqref{sch.fl airline}{\coqdocprojection{fl\_airline}} \coqdocvariable{fl}\coqdocnotation{)} \coqdocnotation{|>} \coqref{sch.travel valid}{\coqdocaxiom{travel\_valid}} \coqdocvariable{fl}\coqexternalref{:type scope:x '->' x}{http://coq.inria.fr/distrib/8.5pl3/stdlib/Coq.Init.Logic}{\coqdocnotation{)}}.\coqdoceol
\coqdocemptyline
\end{coqdoccode}
travels\_consistency is a properties that is verified by the smart
  contract itself. No authority needs to claimed it as is it
  verifiable. \begin{coqdoccode}
\coqdocemptyline
\coqdocnoindent
\coqdockw{Fixpoint} \coqdef{sch.travels consistency}{travels\_consistency}{\coqdocdefinition{travels\_consistency}} (\coqdocvar{tvls} : \coqref{sch.travel itinerary}{\coqdocdefinition{travel\_itinerary}}) (\coqdocvar{tfrom} \coqdocvar{tto}:\coqdocdefinition{cyber.time}):\coqdockw{Prop}:=\coqdoceol
\coqdocnoindent
\coqdockw{match} \coqdocvariable{tvls} \coqdockw{with} \coqdoceol
\coqdocnoindent
\ensuremath{|} \coqexternalref{nil}{http://coq.inria.fr/distrib/8.5pl3/stdlib/Coq.Init.Datatypes}{\coqdocconstructor{nil}} \ensuremath{\Rightarrow} \coqexternalref{True}{http://coq.inria.fr/distrib/8.5pl3/stdlib/Coq.Init.Logic}{\coqdocinductive{True}} \coqdoceol
\coqdocnoindent
\ensuremath{|} \coqdocvar{a}\coqexternalref{:list scope:x '::' x}{http://coq.inria.fr/distrib/8.5pl3/stdlib/Coq.Init.Datatypes}{\coqdocnotation{::}}\coqdocvar{m} \ensuremath{\Rightarrow} \coqexternalref{:nat scope:x '<' x}{http://coq.inria.fr/distrib/8.5pl3/stdlib/Coq.Init.Peano}{\coqdocnotation{(}}\coqexternalref{snd}{http://coq.inria.fr/distrib/8.5pl3/stdlib/Coq.Init.Datatypes}{\coqdocdefinition{snd}} (\coqref{sch.fl departure}{\coqdocprojection{fl\_departure}} \coqdocvar{a})\coqexternalref{:nat scope:x '<' x}{http://coq.inria.fr/distrib/8.5pl3/stdlib/Coq.Init.Peano}{\coqdocnotation{)}} \coqexternalref{:nat scope:x '<' x}{http://coq.inria.fr/distrib/8.5pl3/stdlib/Coq.Init.Peano}{\coqdocnotation{<}} \coqexternalref{:nat scope:x '<' x}{http://coq.inria.fr/distrib/8.5pl3/stdlib/Coq.Init.Peano}{\coqdocnotation{(}}\coqexternalref{snd}{http://coq.inria.fr/distrib/8.5pl3/stdlib/Coq.Init.Datatypes}{\coqdocdefinition{snd}} (\coqref{sch.fl arrival}{\coqdocprojection{fl\_arrival}} \coqdocvar{a})\coqexternalref{:nat scope:x '<' x}{http://coq.inria.fr/distrib/8.5pl3/stdlib/Coq.Init.Peano}{\coqdocnotation{)}} \coqexternalref{:type scope:x '/x5C' x}{http://coq.inria.fr/distrib/8.5pl3/stdlib/Coq.Init.Logic}{\coqdocnotation{\ensuremath{\land}}}\coqdoceol
\coqdocindent{4.00em}
\coqexternalref{:type scope:x '/x5C' x}{http://coq.inria.fr/distrib/8.5pl3/stdlib/Coq.Init.Logic}{\coqdocnotation{(}}\coqdockw{match} \coqdocvar{m} \coqdockw{with} \coqdoceol
\coqdocindent{4.50em}
\ensuremath{|} \coqexternalref{nil}{http://coq.inria.fr/distrib/8.5pl3/stdlib/Coq.Init.Datatypes}{\coqdocconstructor{nil}} \ensuremath{\Rightarrow} \coqexternalref{:type scope:x '=' x}{http://coq.inria.fr/distrib/8.5pl3/stdlib/Coq.Init.Logic}{\coqdocnotation{(}}\coqexternalref{snd}{http://coq.inria.fr/distrib/8.5pl3/stdlib/Coq.Init.Datatypes}{\coqdocdefinition{snd}} (\coqref{sch.fl departure}{\coqdocprojection{fl\_departure}} \coqdocvar{a})\coqexternalref{:type scope:x '=' x}{http://coq.inria.fr/distrib/8.5pl3/stdlib/Coq.Init.Logic}{\coqdocnotation{)}} \coqexternalref{:type scope:x '=' x}{http://coq.inria.fr/distrib/8.5pl3/stdlib/Coq.Init.Logic}{\coqdocnotation{=}} \coqdocvariable{tfrom} \coqexternalref{:type scope:x '/x5C' x}{http://coq.inria.fr/distrib/8.5pl3/stdlib/Coq.Init.Logic}{\coqdocnotation{\ensuremath{\land}}} \coqexternalref{:type scope:x '=' x}{http://coq.inria.fr/distrib/8.5pl3/stdlib/Coq.Init.Logic}{\coqdocnotation{(}}\coqexternalref{snd}{http://coq.inria.fr/distrib/8.5pl3/stdlib/Coq.Init.Datatypes}{\coqdocdefinition{snd}} (\coqref{sch.fl arrival}{\coqdocprojection{fl\_arrival}} \coqdocvar{a})\coqexternalref{:type scope:x '=' x}{http://coq.inria.fr/distrib/8.5pl3/stdlib/Coq.Init.Logic}{\coqdocnotation{)}} \coqexternalref{:type scope:x '=' x}{http://coq.inria.fr/distrib/8.5pl3/stdlib/Coq.Init.Logic}{\coqdocnotation{=}} \coqdocvariable{tto}\coqdoceol
\coqdocindent{4.50em}
\ensuremath{|} \coqdocvar{m} \ensuremath{\Rightarrow} \coqexternalref{:type scope:x '=' x}{http://coq.inria.fr/distrib/8.5pl3/stdlib/Coq.Init.Logic}{\coqdocnotation{(}}\coqexternalref{snd}{http://coq.inria.fr/distrib/8.5pl3/stdlib/Coq.Init.Datatypes}{\coqdocdefinition{snd}} (\coqref{sch.fl departure}{\coqdocprojection{fl\_departure}} \coqdocvar{a})\coqexternalref{:type scope:x '=' x}{http://coq.inria.fr/distrib/8.5pl3/stdlib/Coq.Init.Logic}{\coqdocnotation{)=}}\coqdocvariable{tfrom} \coqexternalref{:type scope:x '/x5C' x}{http://coq.inria.fr/distrib/8.5pl3/stdlib/Coq.Init.Logic}{\coqdocnotation{\ensuremath{\land}}} \coqexternalref{:type scope:x '/x5C' x}{http://coq.inria.fr/distrib/8.5pl3/stdlib/Coq.Init.Logic}{\coqdocnotation{(}}\coqref{sch.travels consistency}{\coqdocdefinition{travels\_consistency}} \coqdocvar{m} (\coqexternalref{snd}{http://coq.inria.fr/distrib/8.5pl3/stdlib/Coq.Init.Datatypes}{\coqdocdefinition{snd}}(\coqref{sch.fl arrival}{\coqdocprojection{fl\_arrival}} \coqdocvar{a})) \coqdocvariable{tto}\coqexternalref{:type scope:x '/x5C' x}{http://coq.inria.fr/distrib/8.5pl3/stdlib/Coq.Init.Logic}{\coqdocnotation{)}}\coqdoceol
\coqdocindent{4.50em}
\coqdockw{end}\coqexternalref{:type scope:x '/x5C' x}{http://coq.inria.fr/distrib/8.5pl3/stdlib/Coq.Init.Logic}{\coqdocnotation{)}}\coqdoceol
\coqdocnoindent
\coqdockw{end}.\coqdoceol
\coqdocemptyline
\end{coqdoccode}
{\it 5- The travel health insurance policy is to be secured,
  covering any medical emergency with hospital care and travel back
  to ones native country due to medical motives. This health
  insurance policy has to cover expenses up to 30,000 euros, the sum
  depending on the residing days, and also it has to be valid in all
  Schengen countries. The health insurance policy must be purchased
  before picking up the visa and if your visa is refused you can
  cancel it!} \begin{coqdoccode}
\coqdocemptyline
\coqdocnoindent
\coqdockw{Variable} \coqdef{sch.travel health}{travel\_health}{\coqdocvariable{travel\_health}} :\coqdockw{Type}.\coqdoceol
\coqdocnoindent
\coqdockw{Variable} \coqdef{sch.travel health of}{travel\_health\_of}{\coqdocvariable{travel\_health\_of}}: \coqref{sch.travel health}{\coqdocaxiom{travel\_health}} \coqexternalref{:type scope:x '->' x}{http://coq.inria.fr/distrib/8.5pl3/stdlib/Coq.Init.Logic}{\coqdocnotation{\ensuremath{\rightarrow}}} \coqdocaxiom{cyber.Authority}.\coqdoceol
\coqdocnoindent
\coqdockw{Variable} \coqdef{sch.travel health emitted by}{travel\_health\_emitted\_by}{\coqdocvariable{travel\_health\_emitted\_by}}: \coqref{sch.travel health}{\coqdocaxiom{travel\_health}} \coqexternalref{:type scope:x '->' x}{http://coq.inria.fr/distrib/8.5pl3/stdlib/Coq.Init.Logic}{\coqdocnotation{\ensuremath{\rightarrow}}} \coqdocaxiom{cyber.Authority}.\coqdoceol
\coqdocnoindent
\coqdockw{Variable} \coqdef{sch.travel health valid}{travel\_health\_valid}{\coqdocvariable{travel\_health\_valid}} : \coqref{sch.travel health}{\coqdocaxiom{travel\_health}} \coqexternalref{:type scope:x '->' x}{http://coq.inria.fr/distrib/8.5pl3/stdlib/Coq.Init.Logic}{\coqdocnotation{\ensuremath{\rightarrow}}} \coqdockw{Prop}.\coqdoceol
\coqdocemptyline
\end{coqdoccode}
{\it 6-Proof of accommodation for the whole duration of the
intended stay in the Schengen area.}\begin{coqdoccode}
\coqdocemptyline
\coqdocnoindent
\coqdockw{Record} \coqdef{sch.accommodation}{accommodation}{\coqdocrecord{accommodation}}:= \coqdef{sch.mkAcc}{mkAcc}{\coqdocconstructor{mkAcc}}\coqdoceol
\coqdocindent{0.50em}
\{ \coqdef{sch.shelter at}{shelter\_at}{\coqdocprojection{shelter\_at}}: \coqdocaxiom{cyber.Authority}; \coqdoceol
\coqdocindent{1.00em}
\coqdef{sch.from}{from}{\coqdocprojection{from}} : \coqdocdefinition{cyber.time}; \coqdoceol
\coqdocindent{1.00em}
\coqdef{sch.to}{to}{\coqdocprojection{to}}:\coqdocdefinition{cyber.time};\coqdoceol
\coqdocindent{0.50em}
\}.\coqdoceol
\coqdocemptyline
\coqdocnoindent
\coqdockw{Definition} \coqdef{sch.accommodations}{accommodations}{\coqdocdefinition{accommodations}}:= \coqexternalref{list}{http://coq.inria.fr/distrib/8.5pl3/stdlib/Coq.Init.Datatypes}{\coqdocinductive{list}} \coqref{sch.accommodation}{\coqdocrecord{accommodation}}.\coqdoceol
\coqdocnoindent
\coqdockw{Variable} \coqdef{sch.accommodation valid}{accommodation\_valid}{\coqdocvariable{accommodation\_valid}}: \coqref{sch.accommodation}{\coqdocrecord{accommodation}} \coqexternalref{:type scope:x '->' x}{http://coq.inria.fr/distrib/8.5pl3/stdlib/Coq.Init.Logic}{\coqdocnotation{\ensuremath{\rightarrow}}} \coqdockw{Prop}.\coqdoceol
\coqdocnoindent
\coqdockw{Definition} \coqdef{sch.accommodations validation}{accommodations\_validation}{\coqdocdefinition{accommodations\_validation}} (\coqdocvar{acc}:\coqref{sch.accommodations}{\coqdocdefinition{accommodations}}):=\coqdoceol
\coqdocnoindent
\coqdockw{\ensuremath{\forall}} \coqdocvar{ac}, \coqexternalref{In}{http://coq.inria.fr/distrib/8.5pl3/stdlib/Coq.Lists.List}{\coqdocdefinition{List.In}} \coqdocvariable{ac} \coqdocvariable{acc} \coqexternalref{:type scope:x '->' x}{http://coq.inria.fr/distrib/8.5pl3/stdlib/Coq.Init.Logic}{\coqdocnotation{\ensuremath{\rightarrow}}} \coqexternalref{:type scope:x '->' x}{http://coq.inria.fr/distrib/8.5pl3/stdlib/Coq.Init.Logic}{\coqdocnotation{(}}\coqdocnotation{(}\coqref{sch.shelter at}{\coqdocprojection{shelter\_at}} \coqdocvariable{ac}\coqdocnotation{)} \coqdocnotation{|>} \coqref{sch.accommodation valid}{\coqdocaxiom{accommodation\_valid}} \coqdocvariable{ac}\coqexternalref{:type scope:x '->' x}{http://coq.inria.fr/distrib/8.5pl3/stdlib/Coq.Init.Logic}{\coqdocnotation{)}}.\coqdoceol
\coqdocemptyline
\coqdocnoindent
\coqdockw{Fixpoint} \coqdef{sch.accommodations consistency}{accommodations\_consistency}{\coqdocdefinition{accommodations\_consistency}} (\coqdocvar{accs}:\coqref{sch.accommodations}{\coqdocdefinition{accommodations}}) (\coqdocvar{tfrom} \coqdocvar{tto}:\coqdocdefinition{cyber.time}):\coqdockw{Prop}:=\coqdoceol
\coqdocnoindent
\coqdockw{match} \coqdocvariable{accs} \coqdockw{with} \coqdoceol
\coqdocnoindent
\ensuremath{|} \coqexternalref{nil}{http://coq.inria.fr/distrib/8.5pl3/stdlib/Coq.Init.Datatypes}{\coqdocconstructor{nil}} \ensuremath{\Rightarrow} \coqexternalref{True}{http://coq.inria.fr/distrib/8.5pl3/stdlib/Coq.Init.Logic}{\coqdocinductive{True}} \coqdoceol
\coqdocnoindent
\ensuremath{|} \coqdocvar{a}\coqexternalref{:list scope:x '::' x}{http://coq.inria.fr/distrib/8.5pl3/stdlib/Coq.Init.Datatypes}{\coqdocnotation{::}}\coqdocvar{m} \ensuremath{\Rightarrow} \coqexternalref{:nat scope:x '<' x}{http://coq.inria.fr/distrib/8.5pl3/stdlib/Coq.Init.Peano}{\coqdocnotation{(}}\coqref{sch.from}{\coqdocprojection{from}} \coqdocvar{a}\coqexternalref{:nat scope:x '<' x}{http://coq.inria.fr/distrib/8.5pl3/stdlib/Coq.Init.Peano}{\coqdocnotation{)}} \coqexternalref{:nat scope:x '<' x}{http://coq.inria.fr/distrib/8.5pl3/stdlib/Coq.Init.Peano}{\coqdocnotation{<}} \coqexternalref{:nat scope:x '<' x}{http://coq.inria.fr/distrib/8.5pl3/stdlib/Coq.Init.Peano}{\coqdocnotation{(}}\coqref{sch.to}{\coqdocprojection{to}} \coqdocvar{a}\coqexternalref{:nat scope:x '<' x}{http://coq.inria.fr/distrib/8.5pl3/stdlib/Coq.Init.Peano}{\coqdocnotation{)}} \coqexternalref{:type scope:x '/x5C' x}{http://coq.inria.fr/distrib/8.5pl3/stdlib/Coq.Init.Logic}{\coqdocnotation{\ensuremath{\land}}}\coqdoceol
\coqdocindent{4.00em}
\coqexternalref{:type scope:x '/x5C' x}{http://coq.inria.fr/distrib/8.5pl3/stdlib/Coq.Init.Logic}{\coqdocnotation{(}}\coqdockw{match} \coqdocvar{m} \coqdockw{with} \coqdoceol
\coqdocindent{4.50em}
\ensuremath{|} \coqexternalref{nil}{http://coq.inria.fr/distrib/8.5pl3/stdlib/Coq.Init.Datatypes}{\coqdocconstructor{nil}} \ensuremath{\Rightarrow} \coqexternalref{:type scope:x '=' x}{http://coq.inria.fr/distrib/8.5pl3/stdlib/Coq.Init.Logic}{\coqdocnotation{(}}\coqref{sch.from}{\coqdocprojection{from}} \coqdocvar{a}\coqexternalref{:type scope:x '=' x}{http://coq.inria.fr/distrib/8.5pl3/stdlib/Coq.Init.Logic}{\coqdocnotation{)}} \coqexternalref{:type scope:x '=' x}{http://coq.inria.fr/distrib/8.5pl3/stdlib/Coq.Init.Logic}{\coqdocnotation{=}} \coqdocvariable{tfrom} \coqexternalref{:type scope:x '/x5C' x}{http://coq.inria.fr/distrib/8.5pl3/stdlib/Coq.Init.Logic}{\coqdocnotation{\ensuremath{\land}}} \coqexternalref{:type scope:x '=' x}{http://coq.inria.fr/distrib/8.5pl3/stdlib/Coq.Init.Logic}{\coqdocnotation{(}}\coqref{sch.to}{\coqdocprojection{to}} \coqdocvar{a}\coqexternalref{:type scope:x '=' x}{http://coq.inria.fr/distrib/8.5pl3/stdlib/Coq.Init.Logic}{\coqdocnotation{)}} \coqexternalref{:type scope:x '=' x}{http://coq.inria.fr/distrib/8.5pl3/stdlib/Coq.Init.Logic}{\coqdocnotation{=}} \coqdocvariable{tto}\coqdoceol
\coqdocindent{4.50em}
\ensuremath{|} \coqdocvar{m} \ensuremath{\Rightarrow} \coqexternalref{:type scope:x '=' x}{http://coq.inria.fr/distrib/8.5pl3/stdlib/Coq.Init.Logic}{\coqdocnotation{(}}\coqref{sch.from}{\coqdocprojection{from}} \coqdocvar{a}\coqexternalref{:type scope:x '=' x}{http://coq.inria.fr/distrib/8.5pl3/stdlib/Coq.Init.Logic}{\coqdocnotation{)=}}\coqdocvariable{tfrom} \coqexternalref{:type scope:x '/x5C' x}{http://coq.inria.fr/distrib/8.5pl3/stdlib/Coq.Init.Logic}{\coqdocnotation{\ensuremath{\land}}} \coqexternalref{:type scope:x '/x5C' x}{http://coq.inria.fr/distrib/8.5pl3/stdlib/Coq.Init.Logic}{\coqdocnotation{(}}\coqref{sch.accommodations consistency}{\coqdocdefinition{accommodations\_consistency}} \coqdocvar{m} (\coqref{sch.to}{\coqdocprojection{to}} \coqdocvar{a}) \coqdocvariable{tto}\coqexternalref{:type scope:x '/x5C' x}{http://coq.inria.fr/distrib/8.5pl3/stdlib/Coq.Init.Logic}{\coqdocnotation{)}}\coqdoceol
\coqdocindent{4.50em}
\coqdockw{end}\coqexternalref{:type scope:x '/x5C' x}{http://coq.inria.fr/distrib/8.5pl3/stdlib/Coq.Init.Logic}{\coqdocnotation{)}}\coqdoceol
\coqdocnoindent
\coqdockw{end}.\coqdoceol
\coqdocemptyline
\end{coqdoccode}
{\it 7- Proof of sufficient means of subsistence during the intended
  stay in the Schengen area. Varies from country to country. To
  complete Supporting document to attest sponsor’s readiness to cover
  your expenses during your stay Proof of prepaid accommodation
  Document about accommodation in private Proof of prepaid
  transport} \begin{coqdoccode}
\coqdocemptyline
\coqdocnoindent
\coqdockw{Inductive} \coqdef{sch.sufficient means}{sufficient\_means}{\coqdocinductive{sufficient\_means}}:=\coqdoceol
\coqdocnoindent
\ensuremath{|} \coqdef{sch.Bank statement}{Bank\_statement}{\coqdocconstructor{Bank\_statement}} : \coqexternalref{:type scope:x '->' x}{http://coq.inria.fr/distrib/8.5pl3/stdlib/Coq.Init.Logic}{\coqdocnotation{(}}\coqdocaxiom{cyber.Authority} \coqexternalref{:type scope:x '->' x}{http://coq.inria.fr/distrib/8.5pl3/stdlib/Coq.Init.Logic}{\coqdocnotation{\ensuremath{\rightarrow}}} \coqdockw{Prop}\coqexternalref{:type scope:x '->' x}{http://coq.inria.fr/distrib/8.5pl3/stdlib/Coq.Init.Logic}{\coqdocnotation{)}} \coqexternalref{:type scope:x '->' x}{http://coq.inria.fr/distrib/8.5pl3/stdlib/Coq.Init.Logic}{\coqdocnotation{\ensuremath{\rightarrow}}} \coqref{sch.sufficient means}{\coqdocinductive{sufficient\_means}}\coqdoceol
\coqdocnoindent
\ensuremath{|} \coqdef{sch.Credit card}{Credit\_card}{\coqdocconstructor{Credit\_card}} : \coqexternalref{:type scope:x '->' x}{http://coq.inria.fr/distrib/8.5pl3/stdlib/Coq.Init.Logic}{\coqdocnotation{(}}\coqdocaxiom{cyber.Authority} \coqexternalref{:type scope:x '->' x}{http://coq.inria.fr/distrib/8.5pl3/stdlib/Coq.Init.Logic}{\coqdocnotation{\ensuremath{\rightarrow}}} \coqexternalref{nat}{http://coq.inria.fr/distrib/8.5pl3/stdlib/Coq.Init.Datatypes}{\coqdocinductive{nat}} \coqexternalref{:type scope:x '->' x}{http://coq.inria.fr/distrib/8.5pl3/stdlib/Coq.Init.Logic}{\coqdocnotation{\ensuremath{\rightarrow}}} \coqdockw{Prop}\coqexternalref{:type scope:x '->' x}{http://coq.inria.fr/distrib/8.5pl3/stdlib/Coq.Init.Logic}{\coqdocnotation{)->}} \coqref{sch.sufficient means}{\coqdocinductive{sufficient\_means}} \coqdoceol
\coqdocnoindent
\ensuremath{|} \coqdef{sch.Cash}{Cash}{\coqdocconstructor{Cash}} : \coqexternalref{:type scope:x '->' x}{http://coq.inria.fr/distrib/8.5pl3/stdlib/Coq.Init.Logic}{\coqdocnotation{(}}\coqexternalref{nat}{http://coq.inria.fr/distrib/8.5pl3/stdlib/Coq.Init.Datatypes}{\coqdocinductive{nat}} \coqexternalref{:type scope:x '->' x}{http://coq.inria.fr/distrib/8.5pl3/stdlib/Coq.Init.Logic}{\coqdocnotation{\ensuremath{\rightarrow}}} \coqdockw{Prop}\coqexternalref{:type scope:x '->' x}{http://coq.inria.fr/distrib/8.5pl3/stdlib/Coq.Init.Logic}{\coqdocnotation{)}} \coqexternalref{:type scope:x '->' x}{http://coq.inria.fr/distrib/8.5pl3/stdlib/Coq.Init.Logic}{\coqdocnotation{\ensuremath{\rightarrow}}} \coqref{sch.sufficient means}{\coqdocinductive{sufficient\_means}}\coqdoceol
\coqdocnoindent
\ensuremath{|} \coqdef{sch.Employment}{Employment}{\coqdocconstructor{Employment}} : \coqexternalref{:type scope:x '->' x}{http://coq.inria.fr/distrib/8.5pl3/stdlib/Coq.Init.Logic}{\coqdocnotation{(}}\coqdocaxiom{cyber.Authority} \coqexternalref{:type scope:x '->' x}{http://coq.inria.fr/distrib/8.5pl3/stdlib/Coq.Init.Logic}{\coqdocnotation{\ensuremath{\rightarrow}}} \coqdockw{Prop}\coqexternalref{:type scope:x '->' x}{http://coq.inria.fr/distrib/8.5pl3/stdlib/Coq.Init.Logic}{\coqdocnotation{)}} \coqexternalref{:type scope:x '->' x}{http://coq.inria.fr/distrib/8.5pl3/stdlib/Coq.Init.Logic}{\coqdocnotation{\ensuremath{\rightarrow}}} \coqref{sch.sufficient means}{\coqdocinductive{sufficient\_means}}.\coqdoceol
\coqdocemptyline
\coqdocnoindent
\coqdockw{Variable} \coqdef{sch.means of sufficiency}{means\_of\_sufficiency}{\coqdocvariable{means\_of\_sufficiency}} : \coqref{sch.country}{\coqdocdefinition{country}} \coqexternalref{:type scope:x '->' x}{http://coq.inria.fr/distrib/8.5pl3/stdlib/Coq.Init.Logic}{\coqdocnotation{\ensuremath{\rightarrow}}} \coqref{sch.sufficient means}{\coqdocinductive{sufficient\_means}} \coqexternalref{:type scope:x '->' x}{http://coq.inria.fr/distrib/8.5pl3/stdlib/Coq.Init.Logic}{\coqdocnotation{\ensuremath{\rightarrow}}} \coqdockw{Prop}.\coqdoceol
\coqdocemptyline
\coqdocnoindent
\coqdockw{Record} \coqdef{sch.Schengen demand}{Schengen\_demand}{\coqdocrecord{Schengen\_demand}} := \coqdef{sch.mkDemand}{mkDemand}{\coqdocconstructor{mkDemand}}\coqdoceol
\coqdocnoindent
\{ \coqdef{sch.form}{form}{\coqdocprojection{form}} : \coqref{sch.schengen form}{\coqdocaxiom{schengen\_form}}; \coqdoceol
\coqdocindent{1.00em}
\coqdef{sch.picture}{picture}{\coqdocprojection{picture}}: \coqref{sch.photo}{\coqdocaxiom{photo}};\coqdoceol
\coqdocindent{1.00em}
\coqdef{sch.pass}{pass}{\coqdocprojection{pass}}: \coqref{sch.passport}{\coqdocaxiom{passport}}; \coqdoceol
\coqdocindent{1.00em}
\coqdef{sch.travels}{travels}{\coqdocprojection{travels}}: \coqref{sch.travel itinerary}{\coqdocdefinition{travel\_itinerary}};\coqdoceol
\coqdocindent{1.00em}
\coqdef{sch.insurance}{insurance}{\coqdocprojection{insurance}}: \coqref{sch.travel health}{\coqdocaxiom{travel\_health}};\coqdoceol
\coqdocindent{1.00em}
\coqdef{sch.lodgings}{lodgings}{\coqdocprojection{lodgings}}: \coqref{sch.accommodations}{\coqdocdefinition{accommodations}};\coqdoceol
\coqdocindent{1.00em}
\coqdef{sch.sufficient}{sufficient}{\coqdocprojection{sufficient}}: \coqref{sch.sufficient means}{\coqdocinductive{sufficient\_means}}; \coqdoceol
\coqdocindent{1.00em}
\coqdef{sch.time stamp}{time\_stamp}{\coqdocprojection{time\_stamp}}:\coqdocdefinition{cyber.time}; \coqdoceol
\coqdocnoindent
\}.\coqdoceol
\coqdocemptyline
\coqdocnoindent
\coqdockw{Definition} \coqdef{sch.schengen demand validation}{schengen\_demand\_validation}{\coqdocdefinition{schengen\_demand\_validation}} (\coqdocvar{requester}:\coqdocaxiom{cyber.Authority})(\coqdocvar{demand\_form}:\coqref{sch.Schengen demand}{\coqdocrecord{Schengen\_demand}}):=\coqdoceol
\coqdocindent{0.50em}
\coqdockw{let} \coqdocvar{demand} := \coqref{sch.form}{\coqdocprojection{form}} (\coqdocvariable{demand\_form}) \coqdoctac{in} \coqdoceol
\coqdocindent{0.50em}
\coqdockw{let} \coqdocvar{C} := \coqref{sch.schengen country}{\coqdocaxiom{schengen\_country}} \coqdocvariable{demand} \coqdoctac{in} \coqdoceol
\coqdocindent{0.50em}
\coqdockw{let} \coqdocvar{Consul}:= \coqref{sch.consulat of}{\coqdocaxiom{consulat\_of}} \coqdocvariable{C} \coqdoctac{in}\coqdoceol
\coqdocindent{0.50em}
\coqdockw{let} \coqdocvar{tfrom} := \coqref{sch.schengen from}{\coqdocaxiom{schengen\_from}} \coqdocvariable{demand} \coqdoctac{in} \coqdoceol
\coqdocindent{1.00em}
\coqdockw{let} \coqdocvar{tto}:= \coqref{sch.schengen to}{\coqdocaxiom{schengen\_to}} \coqdocvariable{demand} \coqdoctac{in} \coqdoceol
\coqdocindent{0.50em}
\coqexternalref{:type scope:x '/x5C' x}{http://coq.inria.fr/distrib/8.5pl3/stdlib/Coq.Init.Logic}{\coqdocnotation{(}}\coqdocvariable{Consul} \coqdocnotation{|>} \coqdocnotation{(}\coqref{sch.schengen form requirement}{\coqdocaxiom{schengen\_form\_requirement}} \coqdocvariable{demand}\coqdocnotation{)}\coqexternalref{:type scope:x '/x5C' x}{http://coq.inria.fr/distrib/8.5pl3/stdlib/Coq.Init.Logic}{\coqdocnotation{)}} \coqexternalref{:type scope:x '/x5C' x}{http://coq.inria.fr/distrib/8.5pl3/stdlib/Coq.Init.Logic}{\coqdocnotation{\ensuremath{\land}}}\coqdoceol
\coqdocindent{0.50em}
\coqexternalref{:type scope:x '/x5C' x}{http://coq.inria.fr/distrib/8.5pl3/stdlib/Coq.Init.Logic}{\coqdocnotation{(}}\coqdocvariable{requester} \coqdocnotation{|>} \coqref{sch.passport photo}{\coqdocaxiom{passport\_photo}} (\coqref{sch.picture}{\coqdocprojection{picture}} \coqdocvariable{demand\_form})\coqexternalref{:type scope:x '/x5C' x}{http://coq.inria.fr/distrib/8.5pl3/stdlib/Coq.Init.Logic}{\coqdocnotation{)}} \coqexternalref{:type scope:x '/x5C' x}{http://coq.inria.fr/distrib/8.5pl3/stdlib/Coq.Init.Logic}{\coqdocnotation{\ensuremath{\land}}}\coqdoceol
\coqdocindent{0.50em}
\coqexternalref{:type scope:x '/x5C' x}{http://coq.inria.fr/distrib/8.5pl3/stdlib/Coq.Init.Logic}{\coqdocnotation{(}}\coqref{sch.travels validation}{\coqdocdefinition{travels\_validation}} (\coqref{sch.travels}{\coqdocprojection{travels}} \coqdocvariable{demand\_form}) \coqexternalref{:type scope:x '/x5C' x}{http://coq.inria.fr/distrib/8.5pl3/stdlib/Coq.Init.Logic}{\coqdocnotation{\ensuremath{\land}}} \coqref{sch.travels consistency}{\coqdocdefinition{travels\_consistency}} (\coqref{sch.travels}{\coqdocprojection{travels}} \coqdocvariable{demand\_form}) \coqdocvariable{tfrom} \coqdocvariable{tto}\coqexternalref{:type scope:x '/x5C' x}{http://coq.inria.fr/distrib/8.5pl3/stdlib/Coq.Init.Logic}{\coqdocnotation{)/\symbol{92}}}\coqdoceol
\coqdocindent{0.50em}
\coqexternalref{:type scope:x '/x5C' x}{http://coq.inria.fr/distrib/8.5pl3/stdlib/Coq.Init.Logic}{\coqdocnotation{(}}\coqdocnotation{(}\coqref{sch.travel health emitted by}{\coqdocaxiom{travel\_health\_emitted\_by}} (\coqref{sch.insurance}{\coqdocprojection{insurance}} (\coqdocvariable{demand\_form}))\coqdocnotation{)} \coqdocnotation{|>} \coqdocnotation{(} \coqref{sch.travel health valid}{\coqdocaxiom{travel\_health\_valid}} (\coqref{sch.insurance}{\coqdocprojection{insurance}} (\coqdocvariable{demand\_form}))\coqdocnotation{)}\coqexternalref{:type scope:x '/x5C' x}{http://coq.inria.fr/distrib/8.5pl3/stdlib/Coq.Init.Logic}{\coqdocnotation{)}} \coqexternalref{:type scope:x '/x5C' x}{http://coq.inria.fr/distrib/8.5pl3/stdlib/Coq.Init.Logic}{\coqdocnotation{\ensuremath{\land}}}\coqdoceol
\coqdocindent{0.50em}
\coqexternalref{:type scope:x '/x5C' x}{http://coq.inria.fr/distrib/8.5pl3/stdlib/Coq.Init.Logic}{\coqdocnotation{((}}\coqref{sch.accommodations validation}{\coqdocdefinition{accommodations\_validation}} (\coqref{sch.lodgings}{\coqdocprojection{lodgings}} \coqdocvariable{demand\_form})\coqexternalref{:type scope:x '/x5C' x}{http://coq.inria.fr/distrib/8.5pl3/stdlib/Coq.Init.Logic}{\coqdocnotation{)/\symbol{92}}} \coqexternalref{:type scope:x '/x5C' x}{http://coq.inria.fr/distrib/8.5pl3/stdlib/Coq.Init.Logic}{\coqdocnotation{(}}\coqref{sch.accommodations consistency}{\coqdocdefinition{accommodations\_consistency}} (\coqref{sch.lodgings}{\coqdocprojection{lodgings}} \coqdocvariable{demand\_form}) \coqdocvariable{tfrom} \coqdocvariable{tto}\coqexternalref{:type scope:x '/x5C' x}{http://coq.inria.fr/distrib/8.5pl3/stdlib/Coq.Init.Logic}{\coqdocnotation{))}} \coqexternalref{:type scope:x '/x5C' x}{http://coq.inria.fr/distrib/8.5pl3/stdlib/Coq.Init.Logic}{\coqdocnotation{\ensuremath{\land}}}\coqdoceol
\coqdocindent{0.50em}
\coqexternalref{:type scope:x '/x5C' x}{http://coq.inria.fr/distrib/8.5pl3/stdlib/Coq.Init.Logic}{\coqdocnotation{(}}(\coqref{sch.means of sufficiency}{\coqdocaxiom{means\_of\_sufficiency}} \coqdocvariable{C}) (\coqref{sch.sufficient}{\coqdocprojection{sufficient}} \coqdocvariable{demand\_form})\coqexternalref{:type scope:x '/x5C' x}{http://coq.inria.fr/distrib/8.5pl3/stdlib/Coq.Init.Logic}{\coqdocnotation{)}}.\coqdoceol
\coqdocemptyline
\end{coqdoccode}

\end{document}